\documentclass[twocolumn,prd,nofootinbib,aps,floats,floatfix,amsmath,amssymb,secnumarabic]{revtex4} %
\usepackage[final]{graphicx}
\pdfoutput=1
\usepackage{amsmath}
\usepackage{bbm}
\usepackage{amsfonts}
\usepackage{amssymb}
\usepackage{latexsym}
\usepackage{graphicx}
\usepackage[english]{babel}
\usepackage{multirow}
\usepackage{float}
\usepackage{url}
\usepackage{slashed}
\usepackage{xcolor}
\usepackage{hyperref}

\def\sfrac#1#2{{\textstyle{#1\over #2}}}
\newcommand{\be}{\begin{equation}}
\newcommand{\ee}{\end{equation}}
\newcommand{\ba}{\begin{array}}
\newcommand{\ea}{\end{array}}
\newcommand{\bea}{\begin{eqnarray}}
\newcommand{\eea}{\end{eqnarray}}
\newcommand{\sss}{\scriptscriptstyle}
\renewcommand{\d}{\mathrm{d}}

\newcommand{\gA}{g_{\sss A}}
\newcommand{\cW}{c_{\sss W}}
\newcommand{\sW}{s_{\sss W}}
\newcommand{\cz}{c_\zeta}
\newcommand{\sz}{s_\zeta}
\newcommand{\sZ}{{\sss Z}}
\newcommand{\sZp}{{\sss Z'}}

\begin{document}

\title{The windows for kinetically mixed $Z'$-mediated dark matter\\
and the galactic center gamma ray excess}
\author{James M. Cline}
\author{Grace Dupuis}
\author{Zuowei Liu}
\affiliation{Department of Physics, McGill University,
3600 Rue University, Montr\'eal, Qu\'ebec, Canada H3A 2T8}
\author{Wei Xue}
\affiliation{INFN, Sezione di Trieste, SISSA, via Bonomea 265, 
34136 Trieste, Italy}

\begin{abstract} 

One of the simplest hidden sectors with signatures in the visible
sector is fermionic dark matter $\chi$ coupled to a $Z'$ gauge boson that has
purely kinetic mixing with the standard model hypercharge.    We
consider the combined constraints from relic density, direct detection
and collider experiments on such models in which the dark matter is either
a Dirac or a Majorana fermion.  We point out sensitivity to details
of the UV completion for the Majorana model.  For kinetic mixing parameter
$\epsilon \le 0.01$, only relic density and direct detection are
relevant, while for larger $\epsilon$, electroweak precision, LHC
dilepton, and missing energy constraints become important.  We identify regions of the
parameter space of $m_\chi$, $m_{Z'}$, dark gauge coupling
and $\epsilon$ that are most promising for discovery through these
experimental probes.  We study the compatibility of the models
with the galactic center gamma ray excess, finding 
agreement at the 2-3$\sigma$ level for the Dirac model.

\end{abstract}
\maketitle

\section{Introduction}

A popular paradigm for dark matter (DM) models is that
there exists a hidden sector \cite{Strassler:2006im,Strassler:2006qa}, including the dark matter particle
and possibly many others, connected to the visible sector
(the standard model, SM) by some weak ``portal'' interactions
\cite{Pospelov:2007mp,ArkaniHamed:2008qn}.
Fermionic dark matter is theoretically attractive because its
mass is protected by chiral symmetry and so does not introduce any
new hierarchies of scale.  It is natural to suppose that it has
some gauge interactions in the hidden sector, of which the simplest
possibility is U(1)$'$ (where the prime distinguishes it from the
SM weak hypercharge).  The portal is gauge kinetic mixing between the
U(1)$'$ field strength $\tilde Z'_{\mu\nu}$ and the SM
hypercharge $Y_{\mu\nu}$ \cite{Holdom:1985ag}:
\be
	-{\epsilon\over 2}\tilde Z'_{\mu\nu} Y^{\mu\nu}.
\label{kmix}
\ee
One is then led to a simple and predictive model where there are only
four essential parameters: $\epsilon$, the U(1)$'$ gauge coupling $g'$,
and the masses $m_\chi$, $m_{Z'}$ of the dark matter $\chi$ and the
U(1)$'$ gauge boson $Z'$.  Although there may be additional particles
at a similar scale, such as a dark Higgs boson to give mass to the 
$Z'$, it is not necessary to assume that they play an essential role,
and it is consistent to consider the model with only four parameters.
These can be constrained to a great extent by assuming a thermal
origin for the DM relic density, and imposing constraints from direct
searches for the DM and collider searches for the $Z'$, as well
as precision electroweak constraints.

The above statements are strictly true when the DM couples vectorially
to the $Z'$.  Another possibility is to have axial vector couplings,
and so we consider both cases
\be
	\tilde Z'_\mu J^\mu_{\tilde Z'} = g'\bar\chi\gamma^\mu \tilde Z'_\mu\chi,\quad 
	\sfrac12 g'\bar\chi\gamma^\mu \gamma_5 \tilde Z'_\mu\chi,
\label{ints}
\ee
where $\chi$ is assumed to be a Dirac particle in the first case,
and Majorana in the second.  This is motivated by the fact that
a Majorana fermion could have couplings
only of the second type (though a Dirac fermion could have couplings
of both types).  
We will refer to these two models as ``Dirac'' and ``Majorana'' dark
matter.  In the Majorana model we are obliged to also consider
dependence upon the mass of the dark Higgs that is responsible for
spontaneous breaking of the U(1)$'$, as will be explained.

This work aims to synthesize the most important
constraints on kinetically mixed $Z'$-mediated dark matter models.
Some aspects of our study are similar to previous ones 
\cite{Dudas:2009uq}-\cite{Pierce:2014spa}, but with the exception
of ref.\ \cite{Hook:2010tw}, these papers study
$Z'$ models that are not just kinetically mixed but have additional
interactions with the standard model.  Ref.\ \cite{Hook:2010tw}
focuses on electroweak precision constraints, while we incorporate
in addition the constraints from relic density, direct detection
and collider physics.  Our analysis
is distinctive in identifying the allowed parameter space in the
well-motivated and economical hidden sector models where the mediation
to the standard model is purely through gauge kinetic mixing. 

We start in sect.\ \ref{models} with a description of the models under
consideration and a discussion of the extent to which they can be
considered complete without reference to physics at higher scales.  In
sect.\  \ref{couplings} the couplings of the $Z'$ to standard model
particles and to the DM are specified, as well as the  visible and
invisible decay widths of the $Z'$. Here we also  briefly discuss
electroweak precision constraints on the finely tuned region of
parameter space where $m_{Z'}\cong m_Z$.   Sect.\  \ref{relic_den}
presents constraints from the relic density assuming that the DM is
thermally produced.  In sect.\ \ref{ddsect} we derive constraints
coming from direct detection, while sect.\  \ref{dilept} deals with
those coming from dilepton searches at the LHC and precision
electroweak studies. Sensitivity of missing energy signals (monojets)
is also discussed.  We synthesize the results in  sect.\
\ref{windows}, giving a summary of the regions of parameter space that
are still allowed, as well as which experimental probes are most
promising for discovery.  In sect.\ \ref{hooperon} we discuss
the potential for these models to address the galactic center gamma
ray excess that has attracted attention recently.
Conclusions are drawn in sect.\ \ref{concl},
and details of cross section calculations are given in the appendices.

\section{Models}
\label{models}

At the phenomenological level, the Dirac DM model is the simplest because
the U(1)$'$ gauge symmetry does not prevent giving a mass to $\chi$
that is unrelated to spontaneous symmetry breaking.  Moreover there
need not be a Higgs field associated with the $Z'$ mass; one can use
the Stueckelberg mechanism \cite{Kors:2004dx} to directly give the
$Z'$ a mass.  Hence it makes sense to consider the Dirac DM  model
as depending upon only the four parameters  $\epsilon$, $g'$,
$m_\chi$, $m_{Z'}$.  One indication of the consistency of this procedure
is the fact that the  DM annihilation cross section for
$\chi\bar\chi\to Z'Z'$ has unitary behavior at large center of mass
energy even if there is only $\chi$ exchange in the $t$-channel, with
no need for Higgs exchange.  The complete theory can be specified
by the kinetic mixing (\ref{kmix}) and the usual terms
\be
	\bar\chi(i\slashed{D}-m_\chi)\chi
	 -\sfrac14 \tilde Z'_{\mu\nu}\tilde Z'^{\mu\nu} 
	- \sfrac12 m^2_{Z'}\tilde Z'_\mu \tilde Z'^\mu
\ee
where $D_\mu = \partial_\mu - ig'\tilde Z'_\mu $ is the covariant derivative.

However for the Majorana DM model, it is not possible to have a bare 
mass term for $\chi$ consistent with the gauge symmetry; the
Stueckelberg mechanism by itself would imply $m_\chi=0$.  To avoid
this, we are obliged to consider spontaneous symmetry breaking, in 
which the dark Higgs boson $h'$ cannot be much heavier than $\chi$
or $Z'$ unless its self-coupling $\lambda'$ is much greater than 
$g'$ or the Yukawa coupling $y'$ that gives rise to $m_\chi = y'\langle
h'\rangle$.  A consequence of this is that the cross section for
$\chi\bar\chi\to Z'Z'$ violates unitarity at high energy unless the
$h'$ exchange diagram is included.  

An ultraviolet complete version of the Majorana model is given by
\bea
	&&\sfrac12\sum_{i=1}^2\left[ \bar\chi_i 
	(i\slashed{\partial}\pm g'\gamma_5\slashed{\tilde Z'})\chi_i
	- y_i \bar\chi_i\left(\phi P_L 
	+ \phi^*P_R\right)\chi_i\right]\nonumber\\
	&& + \left|(\partial_\mu -2ig'\tilde Z'_\mu)\phi\right|^2 - V(\phi)
\eea
where the two Majorana fermions have charge $\pm g'\gamma_5$
to allow for anomaly cancellation, the scalar has charge
$2g'$, and $P_{L,R} = \sfrac12(1\mp \gamma_5)$.  A bare Dirac mass
term $\bar\chi_1\chi_2$ can be forbidden by the discrete symmetry
$\chi_1\to\chi_1$, $\chi_2\to -\chi_2$.  Then after spontaneous
symmetry breaking, we can consider the
lighter of the two mass eigenstates $\chi_{1,2}$ to be the principal
dark matter particle, while the heavier one (also stable) is subdominant, as we will
verify when computing the relic density.    This justifies the neglect
of the extra DM component in our treatment.

\begin{figure*}[t]
\includegraphics[width=2\columnwidth]{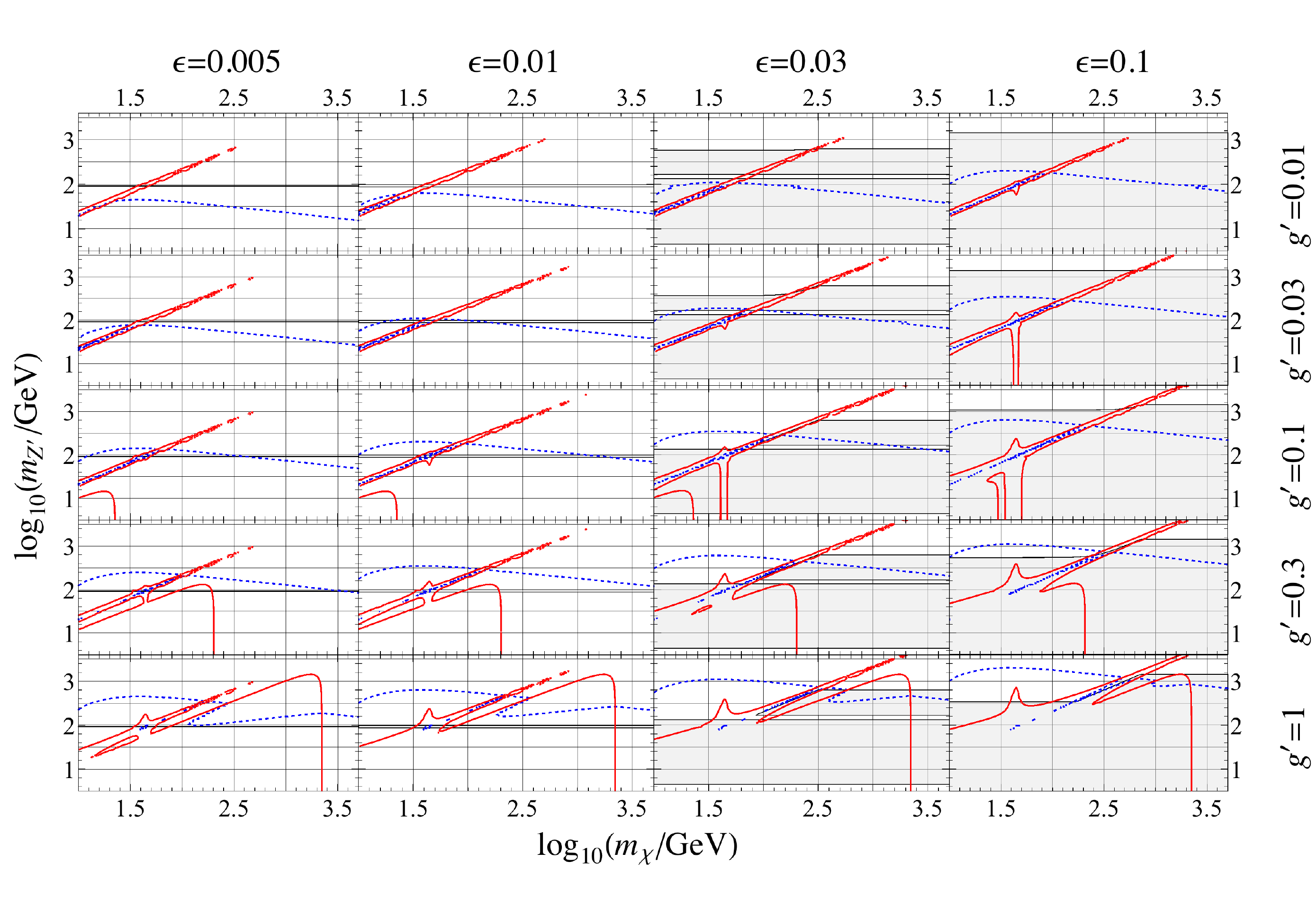}
\caption{Solid (red) curves: relic density contours for the 
Dirac model in the $m_\chi$-$m_{Z'}$ plane.  $\epsilon$ varies
from $0.005$ to $0.1$ (left to right) while $g'$ varies from 
$0.01$ to $1$ (top to bottom).
Dashed (blue) curves denote LUX direct detection upper limits
on $m_{Z'}$.  Shaded regions are excluded by ATLAS dilepton searches
and precision electroweak constraints.}
\label{fig:dirac_relic}
\end{figure*}

\section{Couplings and decays of $Z'$}
\label{couplings}
The couplings of the $Z'$ to standard model particles, via
kinetic mixing, determine the visible contributions to
the width of the $Z'$ and the DM 
annihilation cross section, while the respective 
processes $Z'\to\chi\bar\chi$ or
$\chi\bar\chi\to Z'Z'$ give the invisible contributions, if they
are kinematically allowed.
We distinguished (using the tilde) the interaction eigenstate $\tilde Z'_\mu$ of the 
U(1)$'$ boson that appears in eqs.\ (\ref{kmix},\ref{ints})
from the corresponding mass eigenstate $Z'_\mu$. 
Assuming that there is no 
mass mixing between $Z$ and $Z'$ other than that 
induced by $\epsilon$,  the interaction Lagrangians for the
physical $Z$ and $Z'$ are given by 
 \cite{Cassel:2009pu}
\bea
	{\cal L}_{\rm int} &=& Z_\mu\left(-\epsilon\cW\sz 
	J^\mu_{\rm em} + (\epsilon\sW\sz+\cz) J^\mu_{Z}
	+  \sz J^\mu_{\tilde Z'}\right)\nonumber\\
&+& Z'_\mu\left(-\epsilon\cW\cz 
	J^\mu_{\rm em} + (\epsilon\sW\cz-\sz) J^\mu_{Z}
	+  \cz J^\mu_{\tilde Z'}\right)
\label{Zpcouplings}
\eea
where $\cW = \cos\theta_W$, $\sW = \sin\theta_W$, 
$\cz = \cos\zeta$, $\sz=\sin\zeta$, and we have 
assumed $\epsilon \ll 1$.  The mass mixing angle $\zeta$ is given by
\be
	\tan(\zeta) = {-\epsilon\sW m^2_Z\over 
	m^2_{Z'}-m^2_Z}
\ee
where $m_Z$ represents the SM prediction for the $Z$ boson mass.

In the $Z'$ models considered here, the predicted value of $m_Z$ gets shifted away from the
SM value by an amount $\delta m^2_Z = (m^2_{Z'}-m^2_Z)\tan^2(2\zeta)$,
which is constrained by precision electroweak data, namely the
deviation $\delta\rho$ in the $\rho$ parameter from its SM prediction
$\rho=1$.  This leads to the constraint
\be
	\left|{m^2_{Z'}-m^2_Z\over m^2_Z}\right| >
{\epsilon\sW\delta\rho\over \delta\rho + 2\epsilon\sW}
\ee
where $|\delta\rho| < 10^{-3}$, conservatively.  The maximum allowed
value of $\tan(\zeta)$ is then of order $\delta\rho/\epsilon$.

In the following we will focus on $\epsilon\ge 0.01$, for which
$\zeta$ must therefore be small.  For $m_{Z'} > m_Z$, it is then often
adequate to approximate $\cz=1$, $\sz=0$. For smaller values of
$\epsilon$ this approximation can break down, but only in a finely-tuned
situation where $m_{Z'}$ is very close to $m_Z$.  We will ignore this
possibility in what follows. There are however a few situations where
it is important to keep track of $\zeta$ more acccurately.  One is
when $m_{Z'} \ll m_Z$. In this regime, $\zeta \to \sW\epsilon$ and the
coefficient $(\epsilon\sW\cz-\sz)$ in (\ref{Zpcouplings}) that couples
$Z'$ to the $Z$ current $J^\mu_{Z}$ is highly suppressed.  We will see
that this leads to a strong suppression of the spin-dependent cross
section for scattering of Majorana DM on nucleons.  A second such
situation is  the annihilation $\chi\chi\to f\bar f$ through the $Z$
in the $s$-channel, where we keep $\sin\zeta\ne 0$ since the smallness
of $\zeta$ can be compensated by the $Z$ being nearly on shell in case
of the accidental degeneracy $m_\chi\cong m_Z/2$, leading to resonant
enhancement of the annihilation cross section.

Parametrizing the couplings of the $Z$ and $Z'$ to SM fermions as
\be	\sum_i\bar\psi_i\left[\slashed{Z}\left( v_{i,\sZ} -a_{i,\sZ}
	\gamma_5\right) + \slashed{Z'}\left( v_{i,\sZp} -a_{i,\sZp}
	\gamma_5\right) \right]\psi_i
\label{SMZ'int}
\ee
from (\ref{Zpcouplings}) we find that 
\bea
	v_{i,\sZ} &=& \cz {e\over 2\cW\sW}(T_{3,i}
	-2\sW^2 Q_i)\nonumber\\
	a_{i,\sZ} &=&\cz {e\over 2\cW\sW}T_{3,i}\nonumber\\
	v_{i,\sZp} &=& -\epsilon\cW\cz e Q_i  + (\epsilon\sW\cz-\sz){e\over 2\cW\sW}(T_{3,i}
	-2\sW^2 Q_i)\nonumber\\
	a_{i,\sZp} &=& (\epsilon\sW\cz-\sz){e\over 2\cW\sW}T_{3,i}
\label{ffcouplings}
\eea
where $Q_i$ is the electric charge and $T_{3,i}$ is the weak isospin.
We have ignored corrections of $O(\epsilon^2)$ here.
If $m_{Z'}\gg m_t$ and $\zeta\ll 1$, we can approximate the width of the $Z'$ decaying
into SM particles as
\bea
	\Gamma_{\rm\sss SM} &=& {m_{Z'}\over 4\pi}
	\left(v_{e,\sZp}^2+v_{\nu,\sZp}^2+a_{e,\sZp}^2 +
	a_{\nu,\sZp}^2\right. \nonumber\\ &+& \left. 
	3(v_{u,\sZp}^2+v_{d,\sZp}^2+a_{u,\sZp}^2 +
	a_{d,\sZp}^2)(1+\alpha_s/\pi)\right)\nonumber\\
	&=& {\epsilon^2\alpha m_{Z'}\over 4 \cW^2}\left(
	3 + \sfrac{11}{3}(1+\alpha_s/\pi)\right)
\label{SMwidth}
\eea
The contribution from the top quark should be corrected by the 
factor $(1+\sfrac{7}{17}x)\sqrt{1-4x}$ where $x=(m_t/m_{Z'})^2$
if $x$ is not negligible.  If $m_{Z'} \ll m_Z$, as explained in the
previous paragraph, we cannot approximate $\zeta\cong 0$ because 
of the suppressed coupling of $Z'$ to $J^\mu_{Z}$ (due to the factor
$\epsilon\sW\cz-\sz$).  In that regime,
$Z'$ couples to SM fermions only through the electromagnetic current,
and we find that $\Gamma_{\rm\sss SM}$ is smaller by the factor $4\cW^4/3$
relative to (\ref{SMwidth}).

The invisible width due to $Z'\to \chi\chi$ is given by
\bea
&&\Gamma_{\rm inv} =  \\
&&{g'^2\cz^2\over 12\pi\,m_{Z'}}\,\left(1-{4m_\chi^2\over m_{Z'}^2}
	\right)^{\!\!1/2}
	\left\{\begin{array}{ll} m_{Z'}^2+2m_\chi^2,&{\rm Dirac}\\
	\sfrac12 m_{Z'}^2-2{m_\chi^2},&{\rm Majorana}	
	\end{array}\right.\nonumber
\label{invwidth}
\eea
assuming that $m_{Z'}>2 m_\chi$.

\begin{figure*}[t]
\includegraphics[width=2\columnwidth]{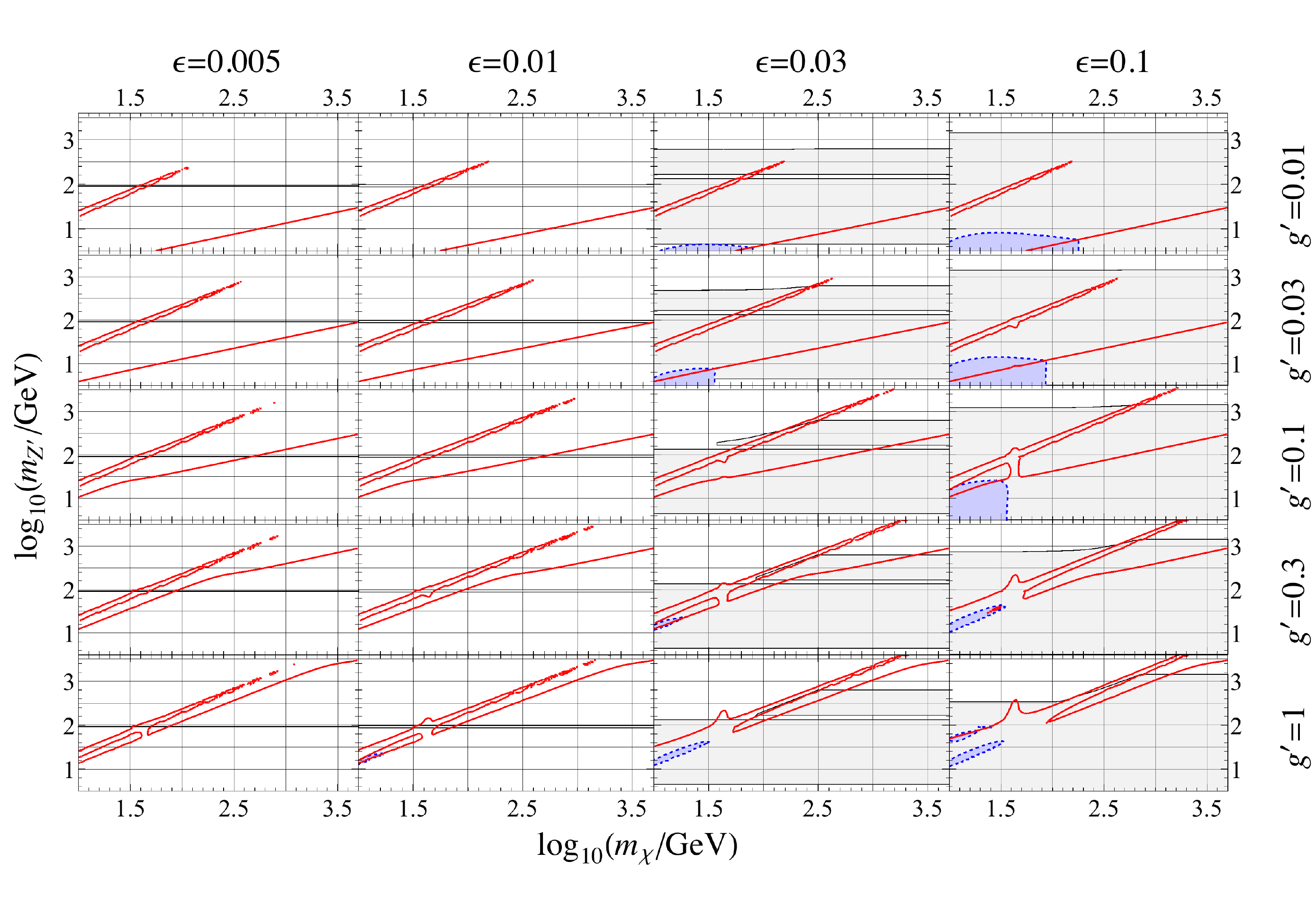}
\caption{Relic density, direct detection and collider constraints
for the Majorana model, as in fig.\ \ref{fig:dirac_relic}. 
Dark (blue) shaded regions bounded by the dashed curves
are excluded by LUX constraint on SI scattering.
The dark Higgs mass is taken to be $m_\phi = m_\chi$.}
\label{fig:maj_relic}
\end{figure*}

\begin{figure}[t]
\vspace{-1cm}
\includegraphics[width=\columnwidth]{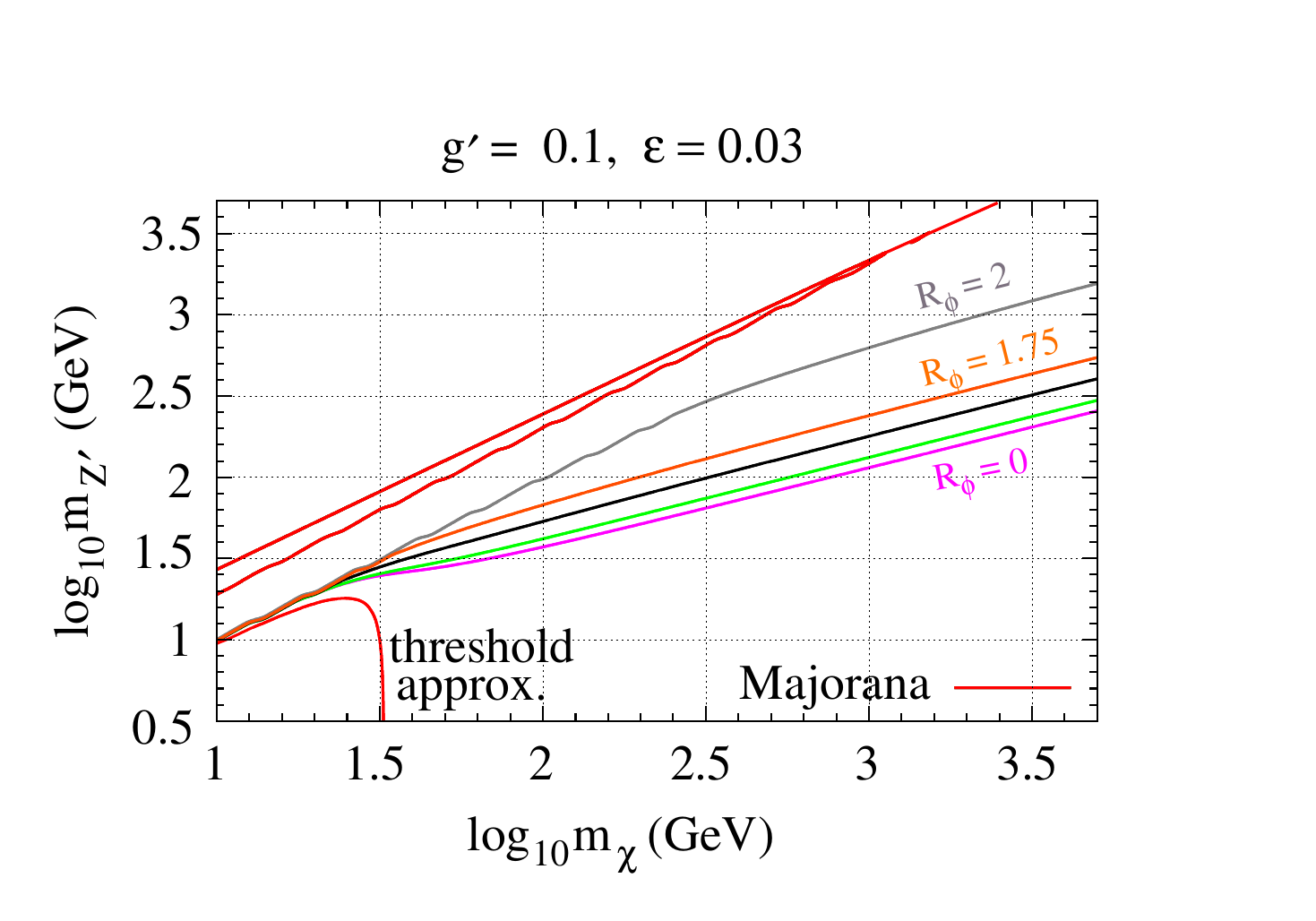}
\caption{Allowed relic density contours for the Majorana model,
with $g'=0.1$, $\epsilon=0.03$. 
Lowest
curve uses the threshold approximation (evaluating cross section 
at $s=4 m_\chi^2$ and omitting the thermal average) for the 
$\chi\chi\to Z'Z'$ contribution to the cross section, rather than the
thermal average of $\sigma v_{\rm rel}$.  Upper curves
show the effect of varying the dark Higgs mass
on the Majorana relic density contours; $R_\phi = m_\phi/m_\chi$ varies between
0 and 2.}
\label{fig:comp}
\end{figure}

\section{Relic density}
\label{relic_den}

There are two potentially important processes for determining the DM
thermal relic density: $\chi \bar\chi \to f\bar f$, where $f$ is any
SM fermion coupling to $Z'$ (the contribution from $W^+W^-$ final
states turns out to be negligible), and $\chi\bar\chi\to Z'Z'$ in the case where
$m_\chi > m_{Z'}$.  The corresponding processes $\chi\bar\chi\to f\bar
f f\bar f$ or $\chi\bar\chi\to f\bar f Z'$, where one or both of the
$Z'$s is off-shell, turn out to give negligible contributions to the
annihilation.  Annihilations into $Z$ bosons can only be important
where $\sz$ is so large that electroweak precision constraints
are violated.  We give details of
the cross section calculations in appendices \ref{ffbar}-\ref{34body}.

To determine the relic density we have solved the full Boltzmann
equation as well as using the accurate approximation described in
ref.\ \cite{Cline:2013gha}.   We find that a faster and accurate
enough method is to compute the thermally averaged cross section
$\langle\sigma v_{\rm rel}\rangle$ at the temperature $T=m_\chi/20$
and compare it to the standard value $(\sigma v)_{\rm th}$ needed
for getting the right relic density.  This quantity has been
accurately determined as a function of $m_\chi$
in ref.\ \cite{Steigman:2012nb}.  Then the
ratio of the $\chi$ relic density to that measured by WMAP7
($\Omega_{\rm CDM}h^2 = 0.112\pm 0.006$) is given by 
\be
	f_{\rm rel} = {g_\chi\over 2}\,{(\sigma v)_{\rm th}
	\over \langle\sigma v_{\rm rel}\rangle}
\label{frel}
\ee
where $g_\chi = 4(2)$ for Dirac (Majorana) DM.  

We display contours 
for $f_{\rm rel}=1$ in the $m_\chi$-$m_{Z'}$ plane for the two models
(Dirac and Majorana DM)
in figures \ref{fig:dirac_relic} and \ref{fig:maj_relic}, for a range
of $g'$ and $\epsilon$.  In nearly all cases, the observational
uncertainty in $\Omega_{\rm CDM}$ does not exceed the widths of the
curves.   There are generally two regions where 
$\langle\sigma v_{\rm rel}\rangle$ has the desired value: one near
$m_{Z'}\cong 2 m_\chi$, where $\chi\chi\to f\bar f$ is resonantly enhanced,
and the second (visible for large enough values of $g'$) where 
$m_{Z'}\cong m_\chi$ so that $\chi\chi\to Z'Z'$ is suppressed by 
lack of phase space.  For Dirac DM, this second branch becomes vertical in the
$m_\chi$-$m_{Z'}$ plane at a sufficiently large value of $m_\chi$, 
beyond which the cross section becomes too small (because of the
suppression from the intermediate $\chi$ propagator in the $t$
channel).  However for Majorana DM, the cross section falls much more
slowly as a function of $m_\chi$, and so the lower branch continues
to large values of $m_\chi$ in this model.  This is related to
the different behavior at large $s$ (the Mandelstam
invariant) in the two models, that was described
in section \ref{models}. 

The slow fall-off of $\sigma v_{\rm rel}$ with $s$  in the Majorana model necessitates doing the full thermal
average to find $\langle\sigma v_{\rm rel}\rangle$, rather than simply
evaluating it at $s=4m_\chi^2$ (the threshold approximation).  In fig.\ \ref{fig:comp} we give an
example (with $g'=0.1$, $\epsilon=0.03$) showing that the latter is a very bad approximation when
$m_\chi$ starts to exceed a certain ($g'$-dependent) value. 
Similarly, the cross section for $\chi\chi\to Z'Z'$ is somewhat
sensitive to the mass of the dark Higgs boson, since its contribution
to the scattering is necessary for getting physically sensible
results.  Whereas we fixed $R_\phi = m_\phi/m_\chi = 1$ in fig.\
\ref{fig:maj_relic}, in fig.\ \ref{fig:comp} we display the
dependence upon $R_\phi$.  There is a marked increase in the cross
section starting at $R_\phi=2$, since the dark Higgs can be produced 
resonantly in that case.  It is worth noting that in this model, the
Yukawa coupling $y_1$ that enters into the scattering matrix element 
is related to the gauge coupling by $y_1/g' = 2m_\chi/m_{Z'}$ 
since both $\chi$ and $Z'$ get their mass from the VEV of $\phi$. 

As a point of consistency for the Majorana model, we require that the
heavier of the two fermions (which was required for anomaly
cancellation) makes a subdominant contribution to the overall relic
density. The contributions to $\sigma v_{\rm rel}$ from the
longitudinal polarizations of the $Z'$ bosons scale as
$m_\chi^2/m_{Z'}^4$, so that the relative abundance of the heavier species is
suppressed by $(m_{\chi_1}/m_{\chi_2})^{2}$.  We numerically verify this
expectation in the high-$m_\chi$, low-$m_{Z'}$ parts of the relic
density contours that are associated with $\chi\chi\to Z'Z'$.

\begin{figure}[t]
\centerline{
\includegraphics[width=\columnwidth,angle=0]{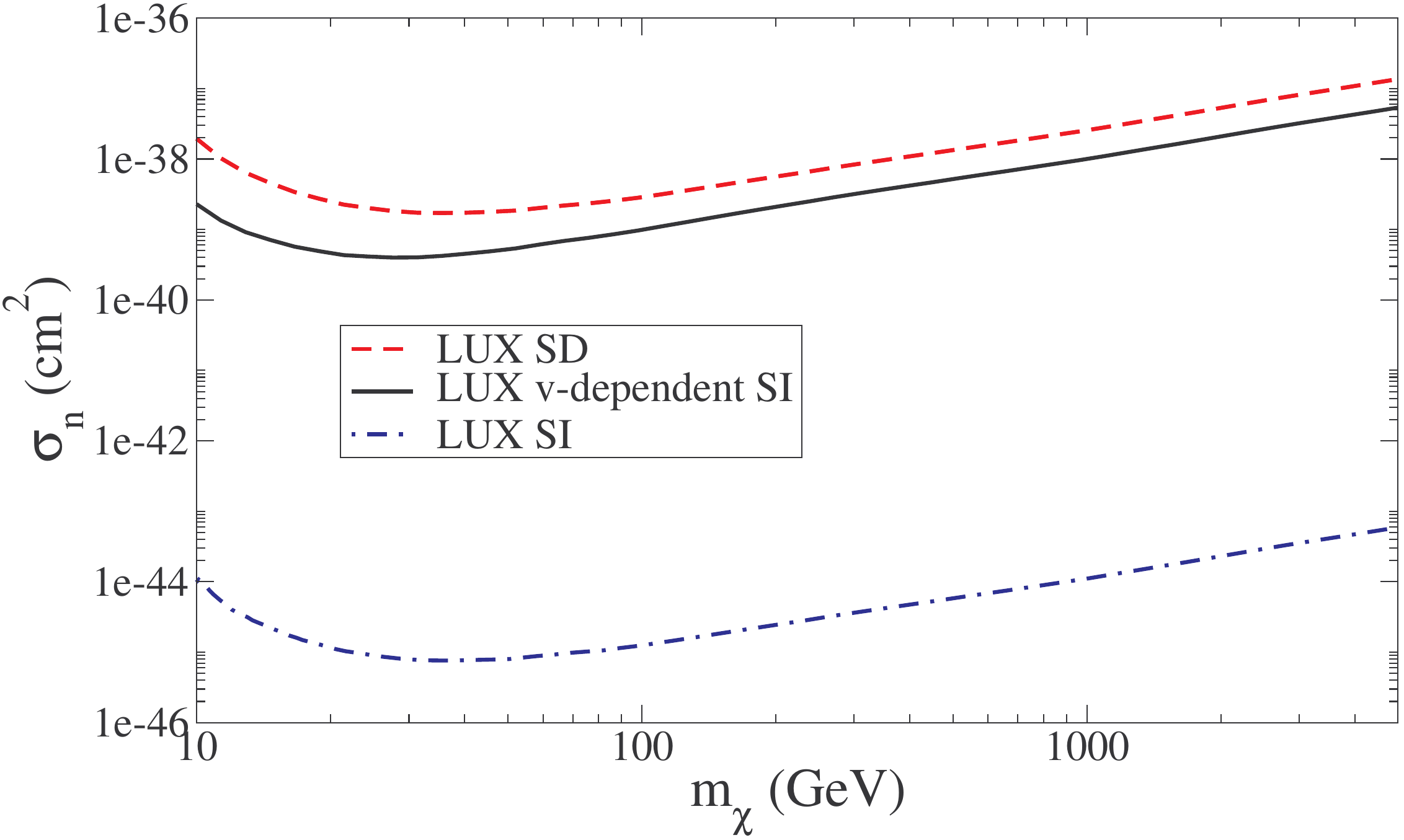}}
\caption{LUX limits on spin-independent, spin-dependent, and
velocity-dependent nucleon cross sections, given respectively by
eqs.\ (\ref{sigSID},\ref{sigSIM},\ref{SDMxsect}).  For the
velocity-dependent case, $\hat\sigma_{\rm\sss SI,M}$ is the constrained
quantity.
}
\label{fig:lux-limits}
\end{figure}

\begin{figure*}[t]
\centerline{
\includegraphics[width=\columnwidth,angle=0]{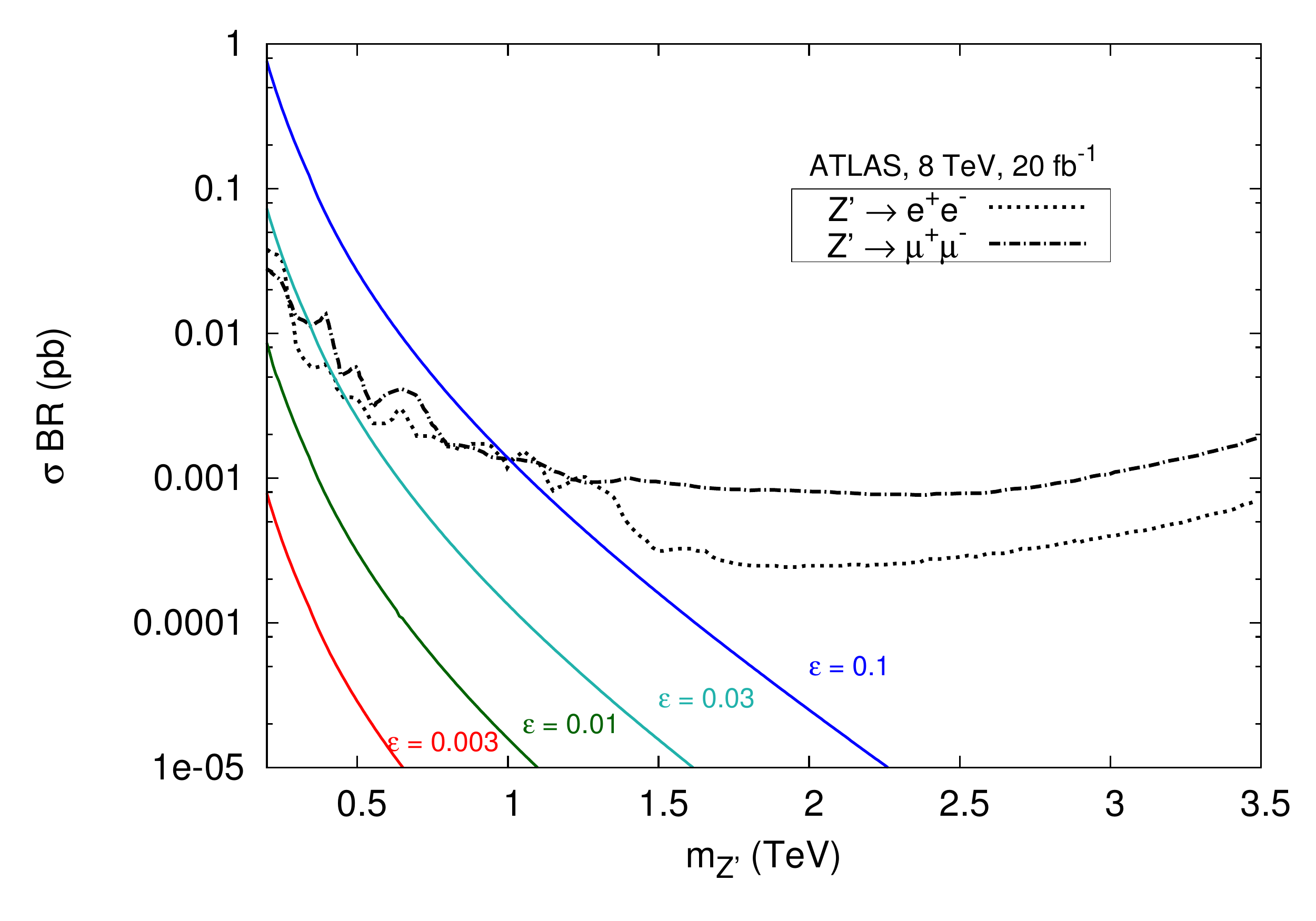}
\includegraphics[width=1.2\columnwidth,angle=0]{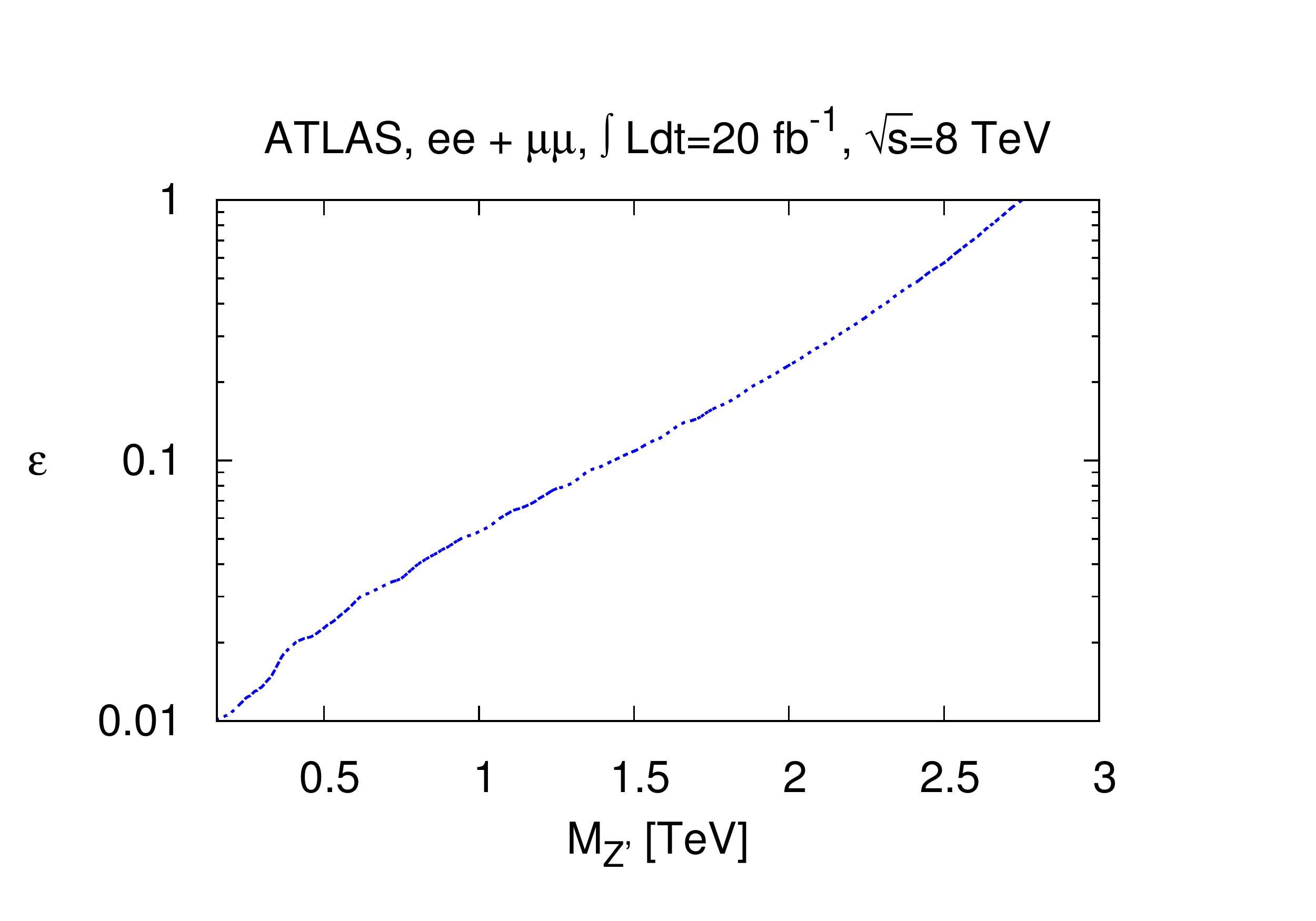}}
\caption{Left: LHC constraints on $Z'$ models from dilepton resonances.
Dashed curves: {ATLAS} limits on $\sigma B(Z'\to l^+ l^-)$.  
Solid curves: Dilepton production cross section as a function of the $Z'$ mass
for several values of $\epsilon$.  Right: Upper bound on $\epsilon$ as a function of $m_{Z'}$ from {ATLAS}
dilepton constraint, using combined electron and muon channels.
}
\label{fig:dilepton}
\end{figure*}

\section{Direct detection}
\label{ddsect}
The cross section for spin-independent (SI) scattering of Dirac DM
$\chi$ to scatter on nucleons at
zero velocity is given by 
\be
	\sigma_{\rm\sss SI,D} = {\mu^2\over \pi}
	\left[\sum_{i=Z,Z'}{Z_N v_{p,i}  + (A_N-Z_N)v_{n,i} \over
	A_N\,m_i^{2}}\, v_{\chi,i}\right]^2
\label{sigSID}
\ee
where $\mu = m_p m_\chi/(m_p+m_\chi)$ is the reduced mass, and
we have averaged over
protons and neutrons to account for coherence, using the charge $Z_N$
and atomic number $A_N$ of the nucleus.  
The vector couplings of the $Z$ and
$Z'$ to the proton and neutron  are  given by 
(\ref{ffcouplings}), which is also valid for
nucleons because of the conserved vector current.
The corresponding couplings to $\chi$ are $v_{\chi,Z}=\sz g'$ and
$v_{\chi,Z'} = \cz g'$.    Numerically, we find that the cross
section is fit to a good approximation by
\be
\sigma_{\rm\sss SI,D} \sim 1.3\times 10^{-30} {\rm cm}^2 (g' \epsilon)^2 
	(m_{Z'}/{\rm GeV})^{-4} 
\ee
for xenon.
However we use the more exact formula (\ref{sigSID}) to obtain the
limits presented below.

For Majorana DM there is a SI contribution to the scattering due
to the vector current at the nucleon, which is suppressed by the
relative velocity, and has different mass dependence:
\bea
	\sigma_{\rm\sss SI,M} &=& {v_{\rm rel}^2}\,{m_n^2 + 2 m_\chi m_n
	+ 3m_\chi^2\over 2(m_\chi + m_n)^2}\, \sigma_{\rm\sss SI,D}
	\nonumber\\
	&\equiv& {v_{\rm rel}^2}\,\hat\sigma_{\rm\sss SI,M}
\label{sigSIM}
\eea
where $m_n$ is the nucleon mass.
There is in addition a spin-dependent (SD) contribution for 
Majorana DM.  We define an effective averaged cross section on
nucleons as  
\bea
  \sigma_{\rm\sss SD,M} &\equiv& \left( \sqrt{ \sigma_{p}} + \sqrt{\sigma_n}
 \right)^2 \\
 &=& {3 \mu^2\over\pi}
\left[\left| \sum_{i} \frac{ a_{p,i} a_{\chi,i} }{ m_i^2} \right|
+ \left| \sum_i \frac{ a_{n,i} a_{\chi,i} }{ m_i^2}\right| \right]^2\nonumber
\label{SDMxsect}
\eea
The axial vector couplings are not simply related to those of
the constituent quarks, instead being given by
\bea
a_{j,Z} &=& e\, {\cz\over{4\cW\sW}}(g_s + 2\gA T_{3,j})\label{axial_couplings}\\
a_{j,Z'} &=& e{\epsilon\sW\cz-\sz\over{4\cW\sW}}(g_s + 2\gA T_{3,j})\nonumber
\eea
where
$g_A = 1.27$ is the axial-vector coupling for neutron decay
and $g_s=0.19$ is the strange quark contribution,
while $a_{\chi,i} = -v_{\chi,i}$.  The actual 
SD cross section $\sigma_N$ on xenon nuclei depends upon a different linear
combination of $a_{p,i}$ and $a_{n,i}$, as described in appendix
 \ref{luxlimits}; the combination $|a_{p,i}|  + |a_{n,i}|$ is just a 
normalization factor in the definition of (\ref{SDMxsect}) that
divides out in the physical $\sigma_N$.  This procedure is consistent
because of the fact that $a_{p,i}/a_{n,i} = -(\gA+g_s)/(\gA-g_s)$
regardless of $i$, a constraint we have imposed when computing the
bound on $\sigma_{\rm\sss SD,M}$.

The LUX direct detection limit can be applied directly to 
$\sigma_{\rm\sss SI,D}$; however we allow for the possibility for $\chi$
to be a subdominant component of the total dark matter by weakening
the constraint according to 
\be
	\sigma_{\rm\sss SI,D} < {\sigma_{\rm\sss SI,LUX}\over f_{\rm rel}}
\label{luxlimit}
\ee
(where ${\sigma_{\rm\sss SI,LUX}}$ is the experimental upper limit)
in regions of parameter space where $f_{\rm rel}< 1$, since the signal
is expected to be reduced by this factor.\footnote{We do
not do so if $f_{\rm rel}>1$ since these cases are ruled out anyway
and they make the graphs harder to read by causing the direct
detection limit to nearly coincide with the relic density curves
in the case of resonantly enhanced annihilation.}
The corresponding constraints on $m_{Z'}$
are shown in fig.\ \ref{fig:dirac_relic} as the dashed (blue) curves.
The use of (\ref{luxlimit}) rather than the more common criterion
$\sigma_{\rm\sss SI,D} < \sigma_{\rm\sss SI,LUX}$ that assumes $f_{\rm rel}=1$
has the virtue that our exclusion curves indicate the true potential
for direct detectability throughout the parameter space, rather than
overestimating it.  

For the velocity- and spin-dependent cross
sections we must determine the limits on $\hat \sigma_{\rm\sss SI,M} \equiv\sigma_{\rm\sss SI,M}/{v_{\rm
rel}^2}$ and $\sigma_{\rm\sss SD,M}$ ourselves, by computing the corresponding
cross sections on the $\mathrm{Xe}^{131}$ nucleus and comparing to the
LUX data.  Details are given in appendix \ref{luxlimits}.
The results are shown in fig.\ \ref{fig:lux-limits}.  

For the Majorana DM model, we find that  the limit on
$\hat\sigma_{\rm\sss SI,M}$ gives more stringent constraints than that on
$\sigma_{\rm\sss SD,M}$, despite the velocity suppression in the
former.\footnote{$\phantom{\!\!\!}$\vspace{-0.05cm}
Stronger limits on SD scattering on protons
in the sun have been obtained by neutrino detection experiments
\cite{Aartsen:2012kia,Avrorin:2014kga}.  These depend upon the efficiency of getting neutrinos from
the decays of final state particles from $\chi\chi$ annihilation.
We have checked that even with the most sensitive channels, the SD
limits obtained are not competitive with the LUX SI limit on our
Majorana DM model.}
  This happens because  the coefficient
$(\epsilon\sW\cz-\sz)$ appearing in  (\ref{axial_couplings}) is
approximately zero for small $m_{Z'}$, making $a_{j,Z'}\cong 0$.  For
heavier $m_{Z'}$, the $Z'$-mediated contribution to the cross section
is suppressed by $1/m_{Z'}^4$.  (Although $(\epsilon\sW\cz-\sz)$ is
also small in the SI cross section for the Dirac model, $v_{p,Z'}$ has
an unsupressed contribution from the $-\cW \cz Q_p$ term.)   The
corresponding limits on $m_{Z'}$ in the Majorana DM model are
given by the dashed (blue) curves in fig.\ \ref{fig:maj_relic},
with dark (blue) shading indicating the excluded regions.

\begin{figure}[t]
\centerline{
\includegraphics[width=\columnwidth,angle=0]{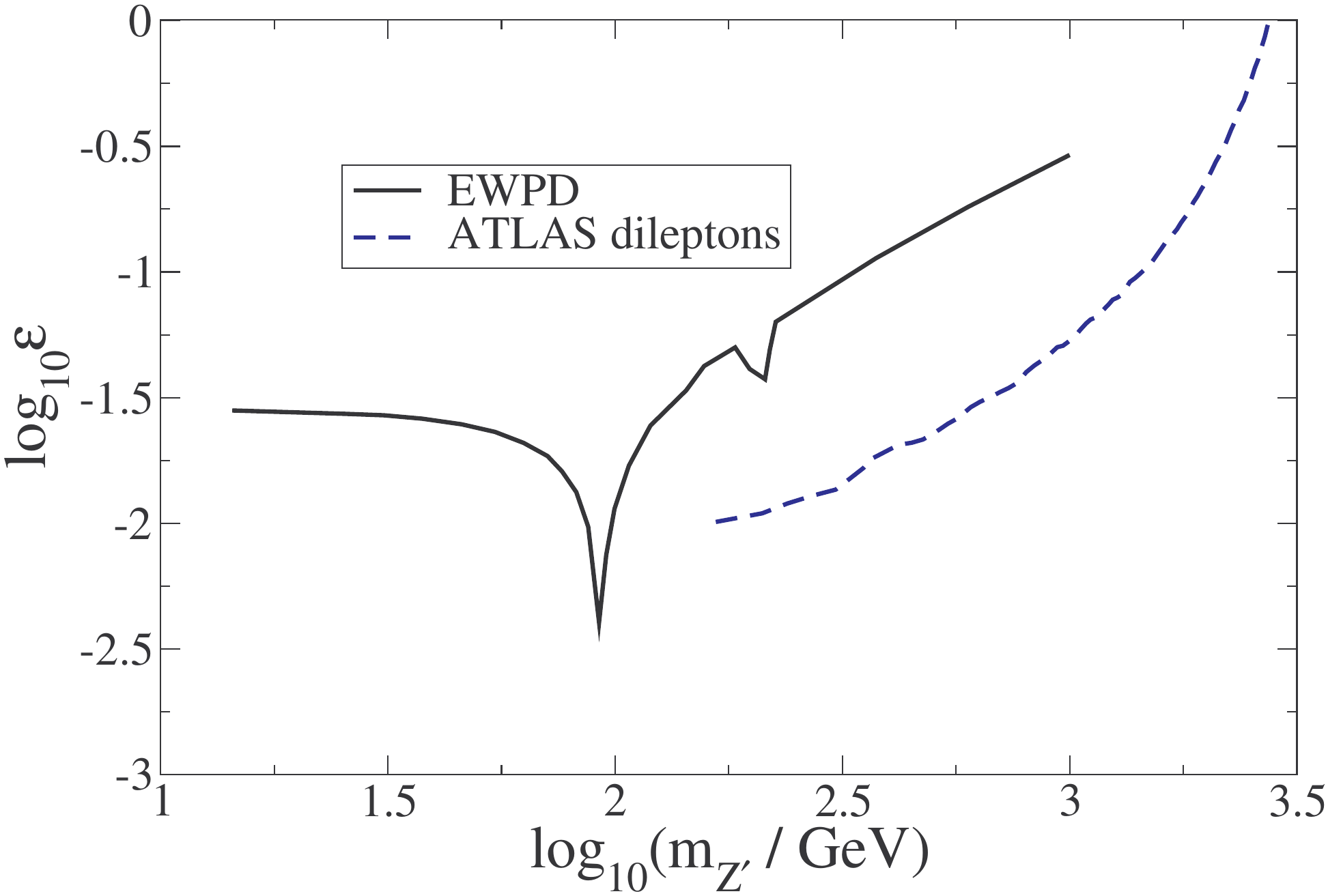}}
\caption{EWPD constraint on ``wide'' $Z'$ from ref.\ 
\cite{Hook:2010tw}, along with our constraint $\epsilon_0$,
eq.\ (\ref{atlas-lim}), from ATLAS dilepton searches.}
\label{fig:eps-lim}
\end{figure}

\begin{figure*}[t]
\centerline{
\includegraphics[width=0.9\columnwidth,angle=0]{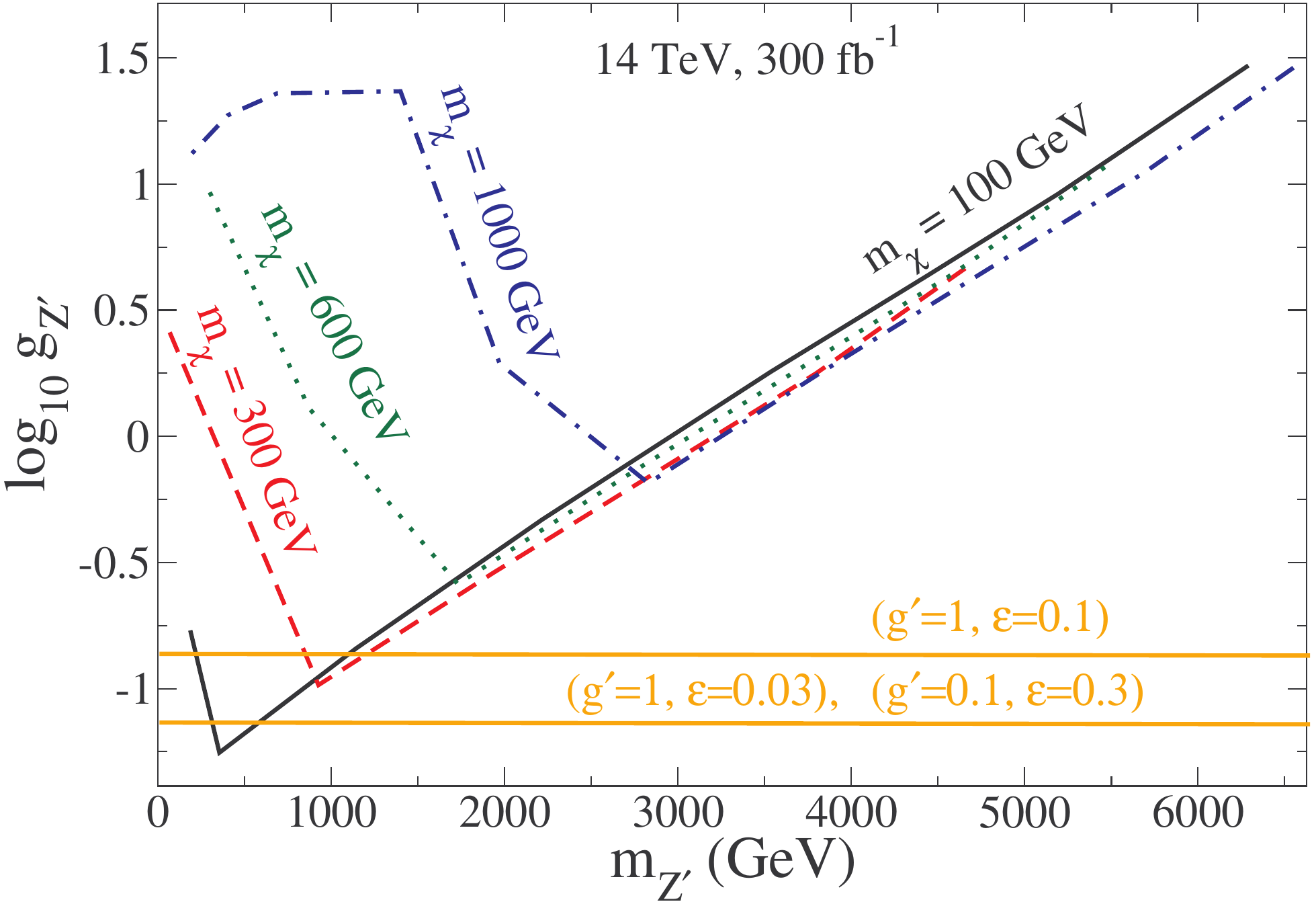}
\includegraphics[width=1.1\columnwidth,angle=0]{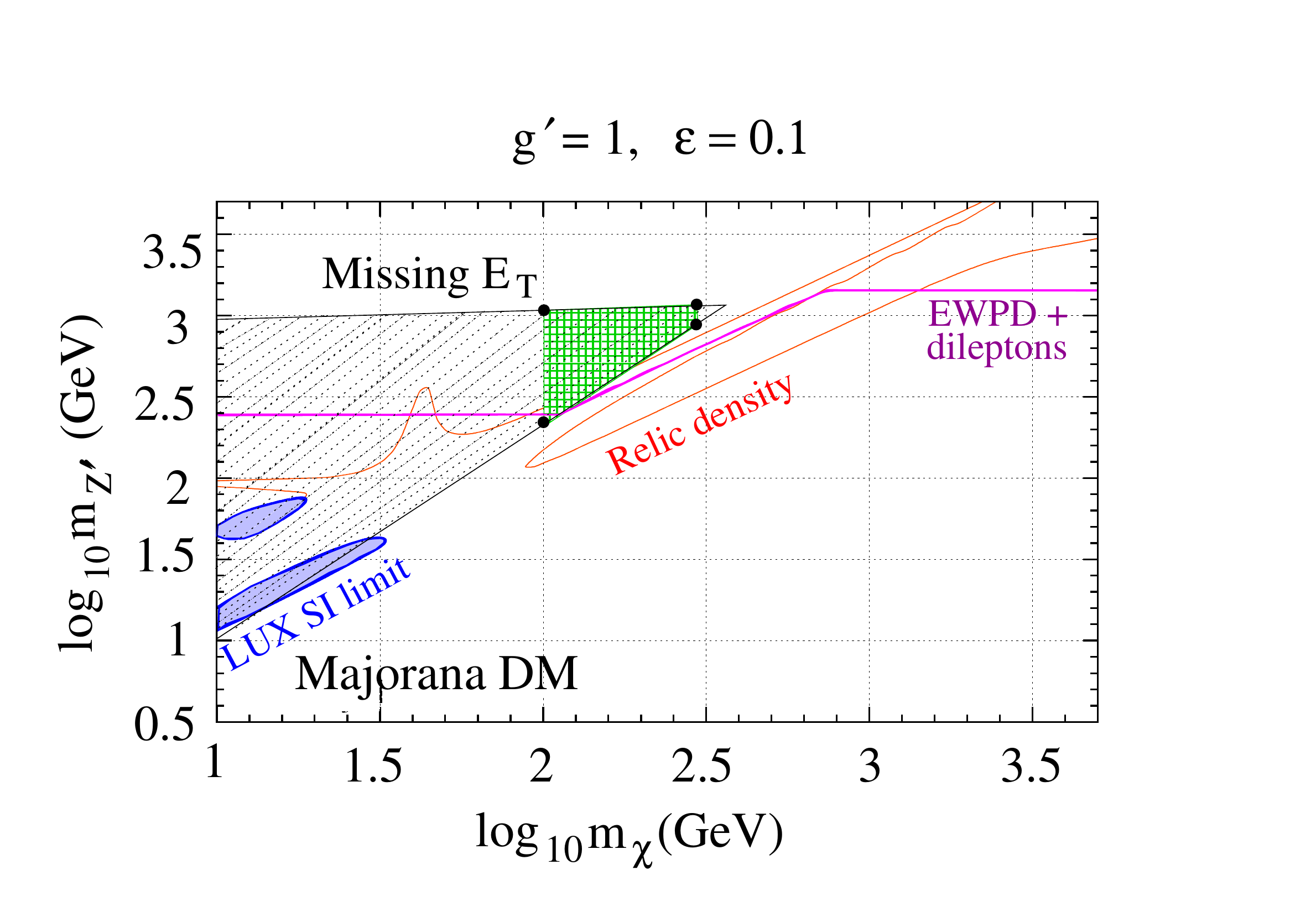}}
\caption{Left: projected LHC upper limits from missing transverse energy 
on $g_{Z'} = (0.175\, g'\epsilon)^{1/2}$
as a function of $m_{Z'}$ for several values of $m_\chi$, 
adapted from ref.\ \cite{Zhou:2013raa}.  Horizontal lines denote
the value of $g_{Z'}$ corresponding to the indicated values of $g'$ 
and $\epsilon$.  Right: plot of the previous regions 
(labeled as ``missing $E_T$'') on 
the $m_\chi$-$m_{Z'}$ plane for the Majorana model with $g'=1$,
$\epsilon= 0.1$, shown as (green) cross-hatched region.  The (black)
hatched region is an extrapolation of results of \cite{Zhou:2013raa}
to lower $m_\chi$.}
\label{fig:monojet}
\end{figure*}

\section{Collider constraints} 
\label{dilept}

There are
constraints on the coupling of $Z'$ to leptons  from the processes
$pp\to Z'\to e^+ e^-,\mu^+\mu^-$  \cite{ATLAS:2013jma}.  These were
derived for other $Z'$ models than the one considered here, so we have
reanalyzed the ATLAS data to constrain the purely kinetically mixed
$Z'$, as described in appendix \ref{app:dilepton}.  In fig.\ 
\ref{fig:dilepton}(a) we show the limits on $\sigma BR$ for 
$Z'\to e^+e^-$ and $Z'\to \mu^+\mu^-$, where $BR$ denotes the
branching ratio for $Z'$ to decay into these final states.  Assuming
that there are no invisible decays, the predicted values of 
$\sigma BR$ for models with a given value of $\epsilon$ are also shown
there.  This allows us to derive the 
upper bound $\epsilon_0(m_Z')$ as a function of $Z'$, assuming that
$Z'$ decays only into SM fermions with the width $\Gamma_{\rm\sss SM}$
given by(\ref{SMwidth}).  The function $\epsilon_0(m_Z')$ is shown
in fig.\ \ref{fig:dilepton}(b).

In general, the above limit must be corrected for the invisible
decays $Z'\to\chi\bar\chi$ through the branching ratio $BR_{\rm\sss SM} =
\Gamma_{\rm\sss SM}/\Gamma_{\rm tot}$, where $\Gamma_{\rm tot} =
\Gamma_{\rm\sss SM} + \Gamma_{\rm inv}$, with $\Gamma_{\rm inv}$ given by
(\ref{invwidth}).  The general constraint is then given by
\be
	\epsilon < {\epsilon_0(m_{Z'})\over (BR_{\rm\sss SM})^{1/2}}
\label{atlas-lim}
\ee
which depends upon both $m_{Z'}$ and $m_\chi$.

The ATLAS limit extends only down to $m_{Z'} = 166$ GeV.  At lower
masses, upper bounds on $\epsilon$ exist from electroweak
precision data (EWPD) constraints \cite{Hook:2010tw}.  We combine these with
(\ref{atlas-lim}) to cover the range down to $m_{Z'} = 10$ GeV.
Generically, the dilepton and EWPD considerations are only relevant
for $\epsilon\gtrsim 0.01$, with slightly more stringent constraints
applying near $m_{Z'} = m_{Z}$ and other narrow mass regions in the
case where $\Gamma_{\rm inv}$ is small.  We adopt the ``wide'' $Z'$
limit of ref.\ \cite{Hook:2010tw}, replotted here in fig.\
\ref{fig:eps-lim}.  For comparison our limit $\epsilon_0$ is also
plotted there.  It should be kept in mind that even though
$\epsilon_0$ is lower than the EWPD limit in the region where they
overlap, EWPD can be more stringent if $BR_{\rm\sss SM}$ is sufficiently
small.

A third collider signal for dark matter models such as those
considered here is missing transverse energy which could occur in the
on-shell production of the $Z'$ if it decays invisibly into
$\chi\bar\chi$.  Initial state radiation from the incoming quarks
could lead to monophotons or monojets.  The ultimate sensitivity of
LHC to $Z'$ models similar to ours has been estimated in ref.\
\cite{Zhou:2013raa}, where projected constraints on the couplings of
the $Z'$ have been computed as a function of $m_{Z'}$ for $m_\chi=100$
and 1000 GeV.  In particular, the effective coupling $g_{Z'} = 
\sqrt{g' g_q}$ is bounded, where $g'$ is the coupling of $Z'$ to
$\chi$, and $g_q$ is its coupling to quarks.  For our purposes,
we take $g_q\cong \epsilon\cW (2e/3)$ corresponding to the up quark
coupling; then $g_{Z'} \cong (0.175\, g'\epsilon)^{1/2}$.  

In fig.\
\ref{fig:monojet}(left), we reproduce the projected limits of 
\cite{Zhou:2013raa}
for the LHC at 14 TeV center-of-mass energy and 300 fb$^{-1}$
integrated luminosity, including rough interpolations to indicate the
limits at intermediate DM masses 300 and 600 GeV.  For comparison,
we draw horizontal lines corresponding to the largest values
of $g'\epsilon = 0.1, 0.03$ considered in 
figs.\ \ref{fig:dirac_relic},\ref{fig:maj_relic}.  We see that the
constraints are somewhat limited; for $g'\epsilon = 0.1$, $m_{Z'}$ is bounded
only for $m_\chi\lesssim 300$ GeV, while for $g'\epsilon = 0.03$
the constraints disappear for $m_\chi\lesssim 100$ GeV.  Nevertheless,
they are complementary to other collider constraints, as shown in
fig.\ \ref{fig:monojet}(right), where we translate the regions
of monojet sensitivity shown previously to display them in the
$m_\chi$-$m_{Z'}$ plane, for the Majorana DM model with $g'=1$, 
$\epsilon=0.1$.  Larger values of $m_{Z'}$ can be probed than
those currently constrained by the dilepton and EWPD studies.
The hatched  region for $m_\chi<100$ GeV is an extrapolation of the
results taken from \cite{Zhou:2013raa}.

\section{Allowed windows}
\label{windows}

In figs.\ \ref{fig:dirac_relic} and \ref{fig:maj_relic} we plot the
contours for the relic density along with upper limits on $m_{Z'}$
from null direct detection searches, and the regions ruled out by
dilepton and EWPD constraints.  As has been noted in previous
literature \cite{Lebedev:2014bba}, the Dirac DM model (fig.\ \ref{fig:dirac_relic}) 
is more highly constrained because of its typically larger cross
section on  nuclei.  For small values of $g'$, the only allowed
regions are the ones where $\chi\chi$ annihilation into SM fermions
is resonantly enhanced due to the accidental tuning of masses
$m_\chi\cong m_{Z'}/2$.  For $g'\epsilon \lesssim 5\times 10^{-5}$,
the direct detection constraint falls below the relic density curve
along $m_\chi\cong m_{Z'}/2$, leaving all such models currently viable.

In the Dirac DM model, only for large values of the U(1)$'$ 
coupling $g'\sim 1$ does the competing channel $\chi\chi\to Z'Z'$
become strong enough to provide an alternative for satisfying both
relic density and direct detection constraints.  This window is
largest for $\epsilon \lesssim 0.01$, below which direct detection and
collider constraints are weakest.  But it survives even for 
$\epsilon$ nearly as large as 0.1, at $m_\chi\cong 1.8$ TeV, $m_{Z'}
\cong 1.4$ TeV. For $\epsilon > 0.03$, the collider/EWPD constraints
become stronger than those from direct detection. 

The Majorana DM model is less constrained because its cross section on
nucleons is either spin-dependent or velocity suppressed.  We found
that the SI (but $v$-dependent) interaction gives the stronger limit.
Even so, it hardly excludes any of the regions favored by the relic
density.   Only for $g'\sim 1$ and $m_\chi\sim m_{Z'}\sim 10$ GeV  is
there significant overlap of the direct detection and relic density
curves. Like in the Dirac model, the relic density can be achieved
either through  $\chi\chi\to f\bar f$ (for $m_\chi\cong m_{Z'}/2$) or
$\chi\chi\to Z'Z'$.   But in contrast, the  
relic density contour  due
to the latter process extends to higher $m_\chi$, due to the 
relatively larger contributions to the annihilation cross section from
the emission of longitudinal gauge bosons. For $\epsilon \gtrsim 0.01$
the collider/EWPD bounds are more important that those for direct
detection, giving the most promising means of discovery.  For $g'\sim
1$, allowed regions with $m_\chi\sim m_{Z'}\sim$ several TeV exist 
even for $\epsilon$ as large as $\sim 0.1$.

\section{Galactic center gamma ray excess}
\label{hooperon}

Evidence from the Fermi Telescope has been found for excess 1-10 GeV 
gamma rays emanating from the galactic center (GC). Although
millisecond pulsars may be a plausible source
\cite{Abazajian:2014fta,Yuan:2014rca}, the possibility of dark matter
annihilation has been vigorously pursued; for a recent discussion with
references see \cite{Daylan:2014rsa}.  Analyses of the data indicate
that 40 GeV dark matter annihilating into $b\bar b$ provide a good fit
to the signal \cite{Abazajian:2014fta}. 

Ref.\ \cite{Izaguirre:2014vva} studied vector and axial-vector mediators in
the $s$-channel, assuming only couplings to dark matter and to $b$
quarks, showing that they are nearly ruled out as an explanation for
the GC excess, by constraints from LUX direct detection and from CMS 
sbottom searches.  On the other hand, 
refs.\ \cite{Martin:2014sxa,Abdullah:2014lla} pointed
out that these constraints are alleviated if $m_{Z'} < m_\chi$ so
that $\chi\chi\to Z'Z'\to 4f$ (where $f$ is a SM fermion) 
can proceed through on-shell $Z'$ bosons
in the GC.  The coupling of $Z'$ to $f\bar f$ can be much smaller in
this case, since the on-shell $Z'$ need only decay eventually into 
SM particles.  Primarily $g'$, $m_\chi$ and $m_{Z'}$ determine the strength
of the GC signal, while the branching ratios of the decays into
different final states affect the shape of the gamma ray spectrum.

\begin{figure}[b]
\centerline{
\includegraphics[width=\columnwidth,angle=0]{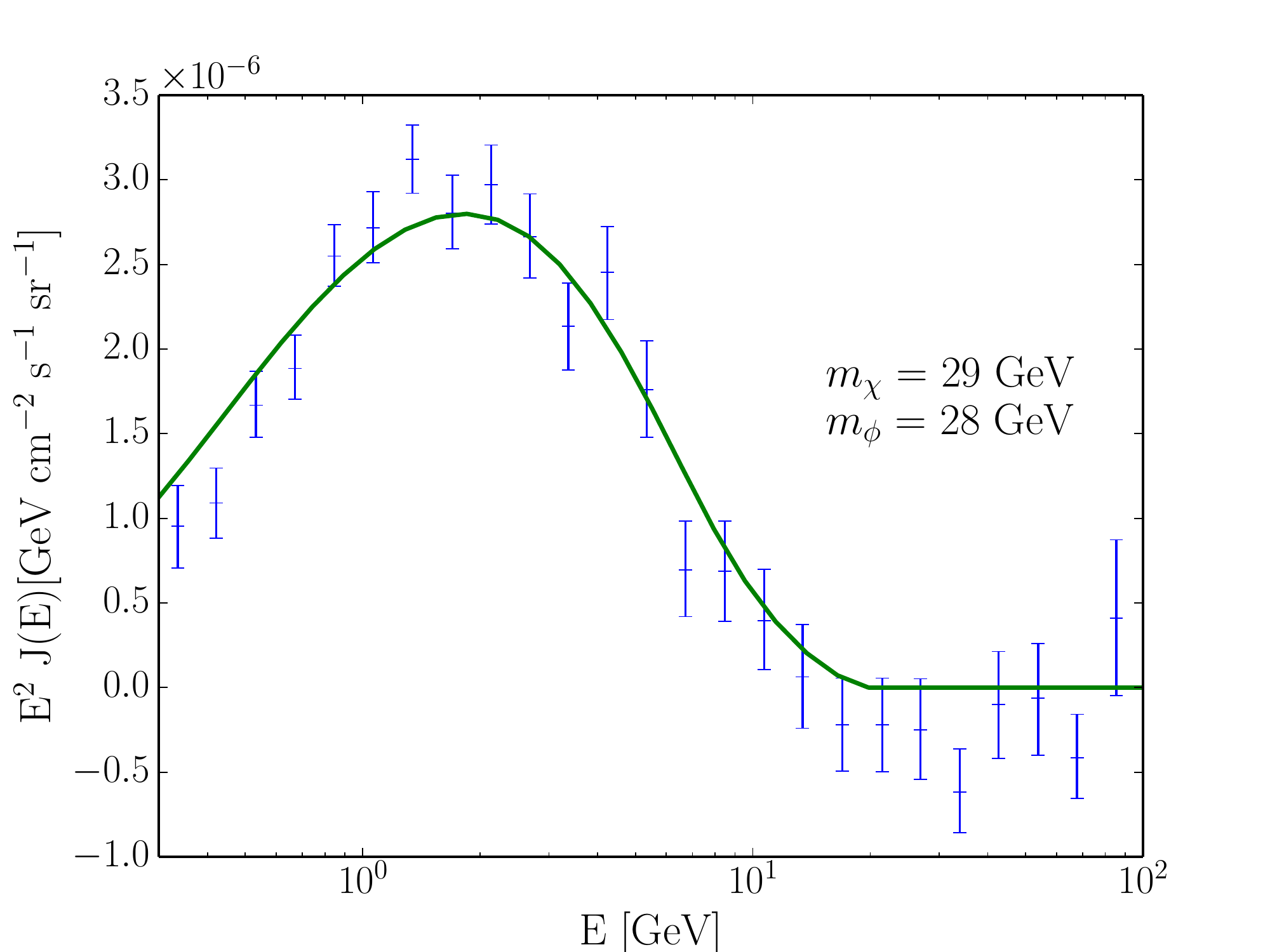}}
\caption{Spectrum of GC gamma ray excess; data are taken from
ref.\ \cite{Daylan:2014rsa}; curve is the best-fit Dirac DM model 
prediction.}
\label{fig:GCspect}
\end{figure}

\begin{figure*}[t]
\centerline{
\includegraphics[width=\columnwidth,angle=0]{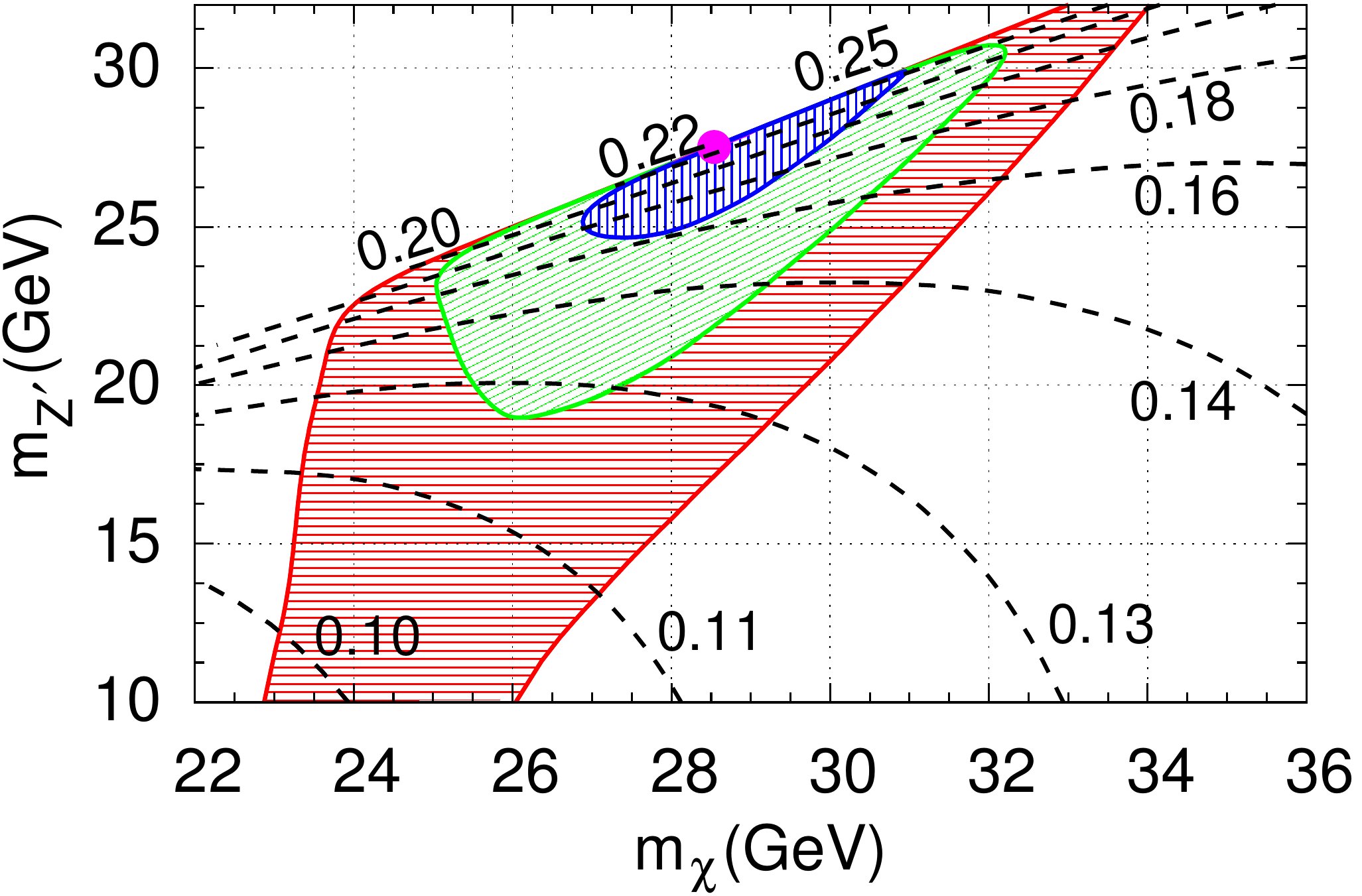}
\includegraphics[width=\columnwidth,angle=0]{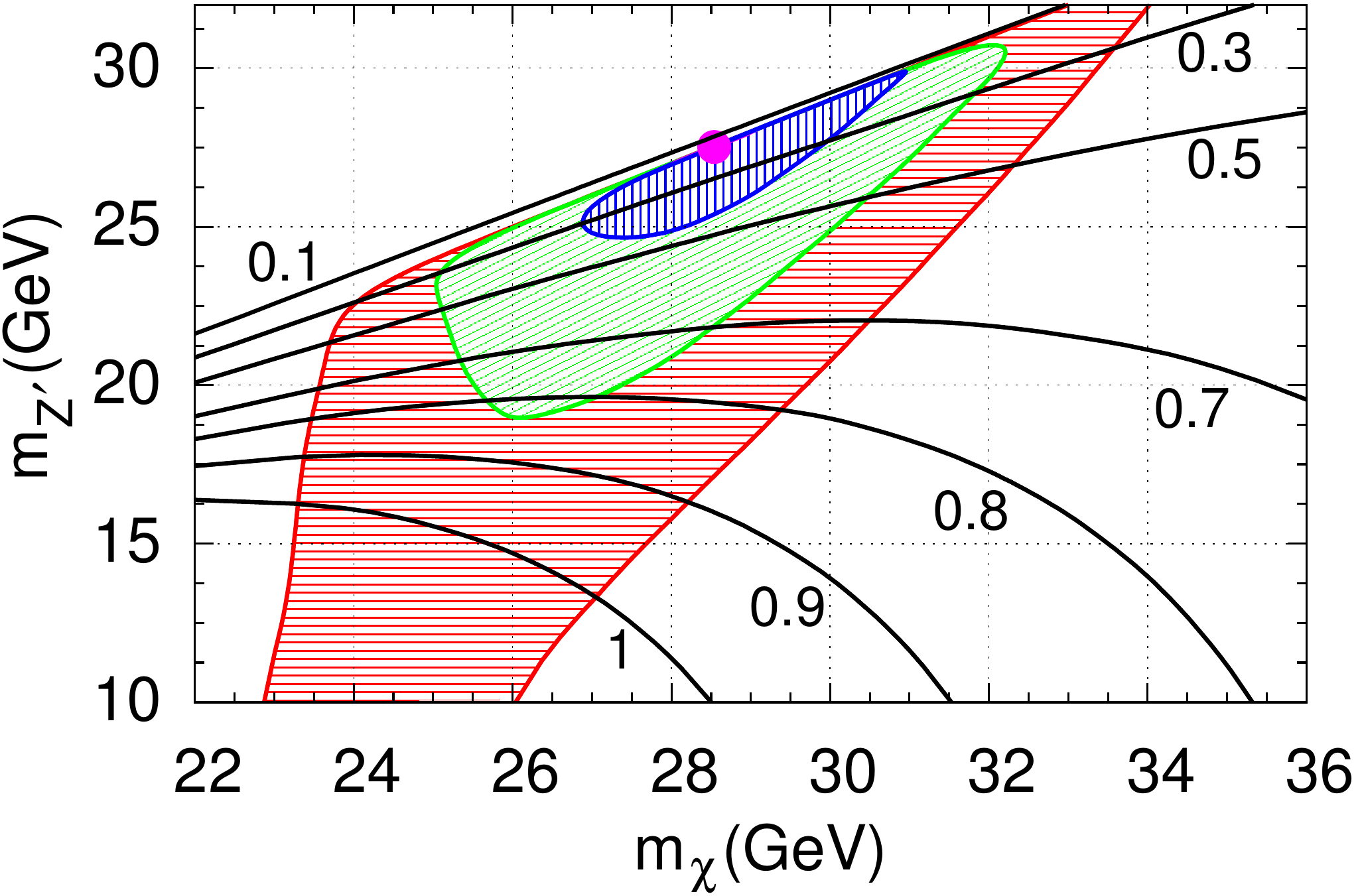}}
\caption{Shaded regions: $1\sigma$, $2\sigma$ and $3\sigma$ confidence
intervals for the galactic center gamma ray excess from cascade
decays of $Z'\to f\bar f$ following $\chi\chi\to Z'Z'$.  
Dot (magenta) indicates best fit point.  Dashed
contours (left) are best-fit values of $g'$.  Solid contours
(right) give the fraction of the relic density $f_{\rm rel}$
for the Dirac DM model,
assuming the $g'$ values indicated on the left.
}
\label{fig:GC}
\end{figure*}

We undertake a similar study here for the case where $Z'$ couples
to the SM through gauge kinetic mixing (this possibility 
was also considered in \cite{Martin:2014sxa}).   Since the models that give
the best fit to the GC excess spectrum have light $Z'$, the couplings
of $Z'$ to fermions are to a good approximation given by the
$-\epsilon\cW\cz e Q_i$ term in (\ref{ffcouplings}), {\it i.e.,} the
$Z'$ couples to their charges.  We have generated the final
photon spectrum using the 
Pythia-based results provided by ref.\ \cite{Cirelli:2010xx},
which mainly considers the processes 
$\chi\chi\to f\bar f$ where each fermion has energy $m_\chi$.
To approximate the effect of 4-body final states, we convolve the photon spectra
from a monoenergetic source with a box distribution, 
\begin{equation}
 \frac{ \mathrm{d} N_\gamma} {\mathrm{d} E_\gamma} 
= \frac{2 }
   {\delta m }
   \int_{(m_\chi - \delta m )/{2} }
^{(m_\chi + \delta m)/{2} }  \d m\,
 \frac{ \mathrm{d} N_\gamma} {\mathrm{d} E_\gamma}(m)  
\end{equation}
where $\delta m\equiv  \sqrt{ m_\chi^2 -m_{Z'}^2}$ and 
$\frac{ \mathrm{d} N_\gamma} {\mathrm{d} E_\gamma}(m)$ is
the spectrum from a 2-body annihilation of particles with mass $m$.
(The factor of 2 accounts for the decays of both $Z'$s.)

To relate the spectrum to the observed gamma-ray flux from the
GC, we use the fact that 
in the galaxy the DM velocity is small, so that 
the zero temperature cross section (\ref{eqB4}) is applicable.
The flux is given by
\begin{equation}
   \frac{\mathrm{d} \Phi} {\mathrm{d} E_{\gamma} \mathrm{d}\Omega } =
     \frac{ r_\odot} { 4 \times 4 \pi} \left( \frac{ \rho_\odot } { m_\chi}
      \right)^2 J \left<\sigma v\right>_0  \frac{\mathrm{d} N_\gamma} { \mathrm{d} E_\gamma}
\end{equation}
where the $J$ factor is the integral along the line of sight
\begin{equation}
 J =  \int_{\rm l.o.s} \frac{\mathrm{d}s } { r_\odot} 
     \left( \frac{\rho_\chi ( r)} { \rho_\odot} \right)^2
\end{equation}
and $\left<\sigma v\right>_0$ is the annihilation cross section at
the kinematic threshold.
We take for the local density at the sun $\rho_\odot  = 0.3\, \mathrm{GeV/cm^3}$ and
$r_\odot = 8.5\, \mathrm{kpc}$.   We compare our theoretical
prediction for the flux to the observed values reported in ref.\ 
\cite{Daylan:2014rsa}, varying $m_\chi$ and $m_{Z'}$ which affect
the shape of the spectrum, and adjusting $g'$ at each
$(m_\chi,\, m_{Z'})$ to obtain the best fit.  We take $\epsilon$
to be negligibly small so that annihilations to $Z'Z'$ dominate
over $f\bar f$ final states and direct detection and collider
constraints are unimportant.   The data and our model's fit to the
spectral shape are shown in fig.\ \ref{fig:GCspect}.

The resulting best-fit regions in the $m_\chi$-$m_{Z'}$ plane are
shown in fig.\ \ref{fig:GC}, along with contours of the corresponding
values of $g'$ (left) and of the relic density fraction for the Dirac
DM model $f_{\rm relic}$ (right).   The best-fit point has
$m_\chi\cong m_{Z'}\cong 28$ GeV, but the 3$\sigma$ confidence region
extends to low values of $m_{Z'}\sim 10$ GeV and $m_\chi\sim 26$ GeV. 
The relic density is too low by a factor of $\sim 6$ at the best-fit
point,  but consistent with the observed value at the lower values of
$m_{Z'}\sim 15\,$GeV.  (For the Majorana DM model, not shown here, the tension between the GC
signal and the relic density is greater, due to the larger thermal
annihilation cross section at the time of freeze-out, even though
at threshold the two models have equal annihilation cross sections.)  The discrepancy between
$f_{\rm rel}$ and the parameters preferred for the GC excess may
be ameliorated by taking into account astrophysical uncertainties
\cite{Abdullah:2014lla}, especially the
possibility of a more concentrated DM halo profile, or accounting for
part of the signal through millisecond pulsar emissions.   Our allowed
regions are similar to those found in ref.\ \cite{Berlin:2014pya}, though somewhat
lower in the masses of $\chi$ and $Z'$.

\section{Conclusions}
\label{concl}

We have systematically studied the constraints from relic density,
direct detection and collider experiments (dilepton production and
electroweak precision data) on a simple dark sector, consisting of
Dirac or Majorana dark matter, connected to the  standard model by a
kinetically mixed massive $Z'$ gauge boson. The Dirac model can be
considered to be UV (ultraviolet) complete, while the Majorana model is somewhat sensitive
to details of the complete theory, such as the mass of the Higgs boson
that spontaneously breaks the U(1)$'$ gauge symmetry, or the presence of
an additional, heavier, subdominant DM component.  

We have shown that the Dirac DM model requires the coincidence
$m_\chi\cong m_{Z'}/2$ to get the right relic density if $\chi$,
and small values of $g'\epsilon$ to evade direct detection, if
$m_\chi\lesssim 300$ GeV.  For heavier DM, there exist allowed models
with larger values of $g'\epsilon$ where $\chi\chi\to Z'Z'$ determines
the relic density, and $\chi$ could be discovered in future searches
for scattering on nuclei or at colliders.  

About the Majorana model, although it has some dependence upon extra
parameters, the qualitative picture is clear: it much more easily
escapes direct detection constraints except for strong couplings
$g'\sim 1$ and small masses $m_\chi\sim m_{Z'}\sim 10$ GeV.  At large
masses, only collider probes are sensitive, and then only for 
relatively large values of the kinetic mixing, $\epsilon \gtrsim
0.01$.  In this regime, models with resonantly enhanced annihilation
($m_\chi\cong m_{Z'}/2$) are more likely to be compatible with the
constraints, unless $g'\gtrsim 0.3$, in which case the more generic
$\chi\chi\to Z'Z'$ branch of the relic-density-allowed regions (with
lower values of $m_{Z'}$) can also be viable.   This region may be
discoverable not only through searches for dileptons but also monojets
in the upcoming run of LHC.

Finally, we studied whether these models can explain the excess 1-10
GeV gamma ray signal from the galactic center found in data from the
Fermi telescope.  There is mild tension between the observed $\gamma$-ray
signal and a thermal origin for the relic density, which is less
severe in the Dirac model, and which would be less significant if the
DM halo profile of the galaxy is more strongly peaked at the center,
or if millisecond pulsars are responsible for part of the observed
excess.   The Dirac DM model is therefore an interesting 
candidate for the GC excess.  

\bigskip

As we were completing this work, ref.\ \cite{Berlin:2014pya} appeared,
which also studied the viability of the light kinetically mixed $Z'$
to explain the galactic center gamma ray excess.

\bigskip
{\bf Acknowledgments.}  We thank Guy Moore for pointing out the
technique discussed in appendix \ref{34body}, Daniel Whiteson
and Ning Zhou 
for helpful information about monojet searches, and Flip Tanedo
and Tim Tait for correspondence about ref.\ \cite{Abdullah:2014lla}.

\appendix
\section{Cross section for $\chi\chi\to f\bar f$}
\label{ffbar}
The cross section for  $\chi\chi\to f\bar f$ is given by
\be
	\sigma v_{\rm rel} = 2\,g'^2\,F(s,\zeta,m_{Z'})\,
 \left\{\begin{array}{ll} \left(s+2\,m_\chi^2\right),& {\rm\
 Dirac}\\ \left( s -4\,m_\chi^2\right),& {\rm\
 Majorana}\end{array} \right.
\ee
where 
\bea
	F &=& \cz^2\, BW(s,Z'){\Gamma_{Z',\rm SM}\over m_{Z'}}
	+ \sz^2\, BW(s,Z){\Gamma_{Z,\rm SM}\over m_{Z}}\nonumber\\
	&+& 2\cz\sz\left[(s-m_Z^2)(s-m_{Z'}^2) + m_Z m_{Z'} \Gamma_Z
	\Gamma_{Z'}\right]\nonumber\\
	&\times& BW(s,Z')\, BW(s,Z) 
	{\Gamma_{\rm mixed}\over (m_{Z'} m_z)^{1/2}}
\eea
and $BW$ stands for the Breit-Wigner distribution
$BW(s,m) = [(s-m^2)^2 + m^2\Gamma^2]^{-1}$ with $\Gamma$ being the
full width, whereas $\Gamma_{X,\rm SM}$ is the partial width for
$X$ to decay into SM fermions.  The ``mixed width'' 
$\Gamma_{\rm mixed}/(m_{Z'} m_z)^{1/2}$
 is defined in analogy to $\Gamma_{Z',\rm SM}/m_{Z'}$
in eq.\ (\ref{SMwidth}), except one should replace $v_{x,Z'}^2\to
v_{x,Z'}v_{x,Z}$ and $a_{x,Z'}^2\to
a_{x,Z'}a_{x,Z}$.


To compute the thermal average of the annihilation cross section,
it is convenient to define dimensionless variables 
$y = s/(4m_\chi^2)$ and $x=m_\chi/T$; the thermal average is then
given by
\be
	\langle\sigma v_{\rm rel}\rangle = 
	{2x\over K_2^2(x)}\int_1^\infty dy\, y\sqrt{y-1}\,K_1(2x\sqrt{y})\, 
	\sigma v_{\rm rel}
\label{therm_avg}
\ee

\section{Cross section for $\chi\chi\to Z'Z'$}
\label{ZpZp}
For the Dirac DM model, $\sigma v_{\rm rel}$ as a function of 
$y=s/(4m_\chi^2)$  and $R = m_{Z'}/m_\chi$ is 
\be
\sigma v_{\rm rel} = {g'^4\over 128\,\pi\, m_\chi^2}
     	\left[{Q_0 Q_1-Q_2 Q_3 Q_l\over  y^{3/2}\sqrt{y-1}\,
Q_d}\right]
\label{svzpzp}
\ee
where
\bea
Q_0 &=& 16\sqrt{y-1}\sqrt{y-R^2}(2y-R^2)\nonumber\\
Q_1 &=& (2 + R^4 + 2y)\nonumber\\
Q_2 &=& -4R^2 + R^4 + 4y\nonumber\\
Q_3 &=& 2(-2 - 2R^2 + R^4 + 4 y + 4 y^2)\nonumber\\
Q_l &=& \log\left[{\left(R^2-2(y+\sqrt{y-1}\sqrt{y-R^2})\right)^4\over
    (-4R^2+R^4 + 4y)^2}\right]\nonumber\\
Q_d &=& (R^2 - 2y)(-4R^2 + R^4 + 4y)
\eea
For the Majorana DM model,  $\sigma v_{\rm rel}$ also takes the form
(\ref{svzpzp}), but with
\bea
Q_1 &=& 16(1-y)(-4R^2 + R^4 + 4y)(3R^4 - 4yR^2 +4y^2)\nonumber\\
   &-&  8(R_\phi^2 - 4y)(-4R^2 + R^4 + 4y)(R^4 - 2yR^2 + 4y^2)
	\nonumber\\ 
    &-& (R_\phi^2 - 4y)^2\left[-R^8 + 2R^4(R^2 + y)\right.\nonumber\\
    &+& \left.8(R^4 - 4yR^2 + 2y^2)\right]\nonumber\\
Q_2 &=& 2(4y-R_\phi^2)(-4R^2 + R^4 + 4y)\nonumber\\
Q_3 &=& 8(R^2 - 2y)(-4R^4 + R^6 + 8yR^2 - 8y^2)\nonumber\\
   &+& (R_\phi^2 - 4y)\left[16y(y-R^2) + 4R^2(R^2 - y)(R^2 +
	4y)\right.\nonumber\\ &-& \left.R^4(R^4 + 4y^2)\right]\nonumber\\
Q_d &=& R^4(R^2 - 2y)(R_\phi^2 - 4y)^2(-4R^2 + R^4 + 4y)
\eea
where $R_\phi = m_\phi/m_\chi$; $\phi$ is the dark sector Higgs
boson that gives rise to $m_{Z'}$.

These cross sections at
threshold are the same for Dirac and Majorana DM in the models
under consideration:
\be
	\sigma v_{\rm rel} = {g'^4\over 16\pi m_\chi^2}\,f(R),
	\quad f(R) \equiv 
	{(1-R^2)^{3/2}\over (1-\sfrac12 R^2)^2}
\label{eqB4}
\ee
However we find that the thermally averaged values can differ
significantly from the threshold values.  This is especially
true for the Majorana model, as described in section 
\ref{relic_den} (see fig.\ \ref{fig:comp}.)

\section{Annihilation into 3 and 4 particles}
\label{34body}
To account for annihilations $\chi\chi\to Z'^* Z'^*$ into off-shell 
$Z's$, without explicitly doing the phase space integrals for the decay
products, one can make the replacement
\bea
	&&2\pi\int {d^{\,4}p\over (2\pi)^4}\, \theta(p_0)\, 
	\delta(p^2-m^2)\quad \to
	\nonumber\\ && \int {d^{\,4}p\over (2\pi)^4}\,
	\theta(p_0)\,\theta(p^2)\, {2\Gamma p_0\over
	(\Gamma p_0)^2 + (p^2-m^2)^2}
\eea	
in the usual invariant phase space integral for each final state
$Z'$, 
where the width is considered as a function of $p_0$.  In the case
that the decay products are approximately massless, $\Gamma = 
\hat\Gamma p_0$.  If we label the energies of the off-shell $Z'$s by
$E_3$ and $E_4$, and their center-of-mass momenta as $p$, 
the cross section $\sigma v_{\rm rel}$ then becomes
an integral over $p$, $E_3$ and $E_4$, with a delta function
$\delta(\sqrt{s} - E_3 - E_4)$.  Rather than using this delta function
to eliminate one of these integrals, it is convenient to save it for
doing the integral over $s$ in the thermal averaging.  The result can
be written as
\bea
\langle\sigma v_{\rm rel}\rangle &=& {x\over 2\pi^3 m_\chi^3
K_2^2(x)} \int_0^\infty dp\,p^2\int_p^\infty dE_3\,
 \int_p^\infty dE_4\,\nonumber\\
&\times&\langle|{\cal M}|^2\rangle \sqrt{y(y-1)}\, K_1(2x\sqrt{y})\,
\theta(y-1)
\nonumber\\
&\times&\prod_{i=3}^4 {\hat\Gamma E_i^2\over (\hat\Gamma E_i^2)^2 + 
	(p_i^2 - m_{Z'}^2)^2}
\label{therm_res}
\eea
where $y=(E_3 + E_4)^2/(2 m_\chi)^2$ and $p_i^2 = E_i^2 - p^2$.

To derive this, start with the Lorentz-invariant expression for 
$v_{\rm rel}$:
\be
	v_{\rm rel} = {2 p_{1,\rm cm}\over E_{1,\rm cm}}
	= 4{\sqrt{(p_1\cdot p_2)^2 - m_\chi^4}\over s}
\ee
Then 
\be
	d\sigma v_{\rm rel} = 
	{(2\pi)^4\over s}\langle|{\cal M}|^2\rangle 
	\, d\Phi_2
\ee
where 
\be
	d\Phi_2 = \delta^{(4)}(p_i)\prod_{3,4}{d^3p_i\over(2\pi)^3 2
	E_i}
\ee
in the usual formulation.  We modify the phase space according to
\be
	{d^3p\over(2\pi)^3 2E} \to 2 {d^4p\over(2\pi)^4} {(\Gamma
	p^0)\,\theta(p^0)\,\theta(p^2)\over (\Gamma p^0)^2 + (p^2-m^2)^2}
\ee
Substitution of the resulting $\sigma v_{\rm rel}$ into 
(\ref{therm_avg}) results in (\ref{therm_res}).

Taking the limit $\hat\Gamma\to 0$ puts the final state $Z'$s on shell
and removes the integrals over $E_i$.  Naively, it would seem valid
to take this limit whenever the energy width of the thermal factor,
which goes like $\exp((2m_\chi-E_3-E_4)/T)$, is bigger than that of the
Breit-Wigner factors.  This is true when $T\gg \hat m_{Z'}$, or 
equivalently if freeze-out happens for $\hat\Gamma 
\ll m_{\chi}/(x_f m_{Z'})$.  In our model, this implies we can put the $Z'$s on shell as
long as 
\be
	{m_{Z'}\over m_\chi} \ll {1\over \epsilon^2}
\label{narrow_app}
\ee	
in which case $\langle\sigma v_{\rm rel}$ does not depend upon
$\epsilon$.  Otherwise it is necessary to do all three integrals and
the result will be suppressed by some power of $\epsilon$.

The above argument misses the cases where only one of the
$Z'$s is on shell, which dominate for some intermediate range of $R$.
However in our numerical study we find that the 3- and 4-body channels
make a small contribution to the total annihilation cross section,
which we therefore ignore. 

\section{LUX limit on SD and velocity-suppressed scattering}
\label{luxlimits}

\begin{figure}[t]
\centerline{
\includegraphics[width=\columnwidth,angle=0]{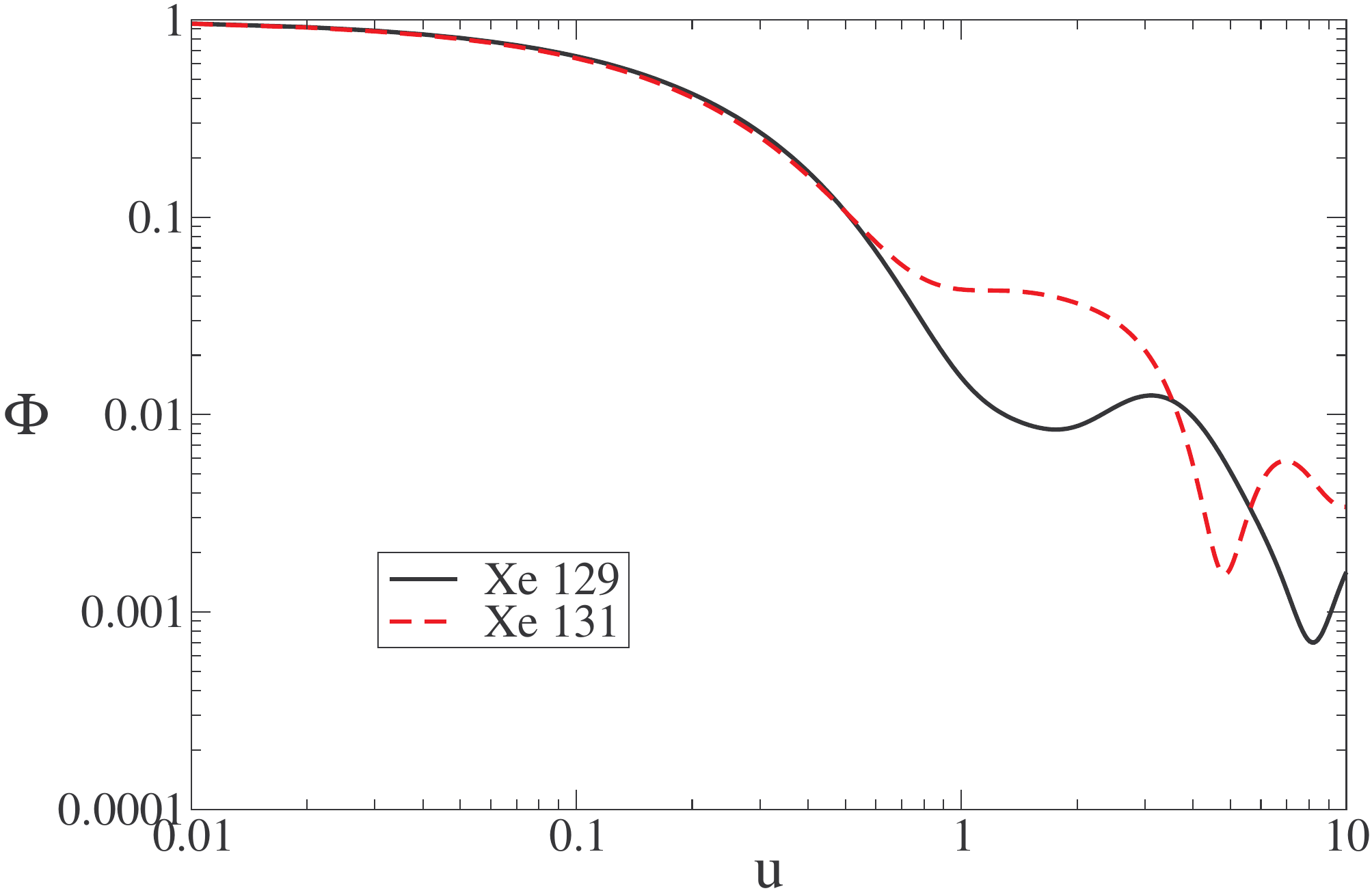}}
\caption{Form factors for spin-dependent scattering on  
$\mathrm{Xe}^{129}$ and $\mathrm{Xe}^{131}$ as a function of 
$u=q^2 b^2/2$ (see text).}
\label{fig:ffact}
\end{figure}

To compute the LUX spin-dependent (SD) scattering limit, the DM recoil rate is given by  
 \begin{equation} 
  \frac{ \d R } { \d E_R} = {\rm Eff} \times{\rm Exp}  \times N_T
  \frac{ \rho_\oplus } { m_{\chi} }
   \int \d^3 v  v f_{\oplus} ( v) \frac{ \d \sigma} { \d E_R} \ ,
\label{dRdER}
\end{equation}
where $N_T$ is the number of targets, $\rho_\oplus = 0.3$ GeV/cm$^3$, and
the Maxwell-Boltzmann velocity distribution $f_\oplus(v)$ is 
assumed.
The exposure ${\rm Exp}$ is (85 live days) $\times$ (118 kg) and the efficiency curve
$\rm{Eff}(E_R)$ is provided by the LUX group.
The DM-nucleus cross section rate gets contributions from
two isotopes weighted by
their abundances $\alpha_i$,
$21.8\%$ for $\mathrm{Xe}^{131}$ and 
$26.2\%$ for $\mathrm{Xe}^{129}$, 
\begin{eqnarray}
   \frac{\d \sigma } { \d E_R}  &=& \sum_{i = \left({Xe^{131},\atop
Xe^{129}}\right)}\!\!\!\! \alpha_i \sigma_{SD}
   \frac{ m_N^2 }{ 2 \mu_N v^2}  
   \frac{ \mu_N^2} { \mu_n^2} \frac{4}{3} \frac{ J+1}{J} 
   \nonumber \\
   &&  \frac{ \left( a_p \left<S_p\right> + a_n \left<S_n\right> \right)^2}
      { \left( |a_p| + |a_n| \right)^2 }  \Phi_i(q) \ .
\end{eqnarray}
We take the
spin matrix elements of the neutron $\left< S_n \right>$ and proton
$\left< S_p \right>$
from table I in 
\cite{Cannoni:2012jq}. 
For $\mathrm{Xe}^{131}$, $J = 3/2$, $\left< S_n \right> = -0.242$, 
$\left< S_p \right> = -0.038$;
for $\mathrm{Xe}^{129}$, $J = 1/2$, $\left< S_n \right> = 0.293$, 
$\left< S_p \right> = 0.046$.  
Since we are considering a wide range of DM masses, 
and at large $m_\chi$
the momentum dependence makes an essential correction to the cross section,
the two nuclear form factors $\Phi_i(q)$ for $\mathrm{Xe}^{131}$ and 
$\mathrm{Xe}^{129}$ are taken into account here \cite{Cannoni:2012jq}.
Following ref.\ \cite{Menendez:2012tm}, we take the form factor for each element to
be
\be
	\Phi_i(q) = {\gA^2 S_{00}^{(i)}(q) + \gA g_s S_{01}^{(i)}(q) + g_s^2
S_{11}^{(i)}(q) \over \gA^2 S_{00}^{(i)}(0) + \gA g_s S_{01}^{(i)}(0) + g_s^2
S_{11}^{(i)}(0)}
\ee
The result is plotted in fig.\ \ref{fig:ffact} as a function of 
$u \equiv q^2 b^2/2$, where $q=\sqrt{2m_N E_R}$ is the momentum
transfer and $b = 2.2853$\,fm (2.2905\, fm) for 
$\mathrm{Xe}^{129}$ ($\mathrm{Xe}^{131}$).

For a given DM model, the predicted number of events is computed by
integrating the recoil rate over the recoil energy from $3~
\mathrm{keV_{nr}}$ to  $38~ \mathrm{keV_{nr}}$.  The upper limit of
the DM cross section is derived by comparing the predicted number of
events with the expected signal events, which ranges from 2.4 to 5.3
for different dark matter masses.

The rate for spin-independent (SI) scattering is also given by
an expression of the form (\ref{dRdER}).  The only difference relative
to standard SI scattering in the case of the Majorana model is
the extra dependence on $v_{\rm rel}^2$ of (\ref{sigSIM}), appearing 
in the phase space integral in (\ref{dRdER}).

\section{Dilepton production cross section}
\label{app:dilepton}
The predicted cross section for dilepton production 
 at the LHC is given by
\bea
	&&\!\!\!\!\!\!\!\!\!\!\!\!\!\!\!\!\!\!\!\!{d\sigma(pp\to l^+ l^-)\over dM } = 
\\ && K {4M\over s} \int_{1}^{\tau} 
	{dx\over x}f_q(\tau)f_{\bar{q}}(\tau^2 / x)
	 \hat{\sigma}(q\bar{q} \to l^+ l^-)\nonumber
\eea
where $M$ is the invariant mass of the lepton pair, $\sqrt{s}=8$ TeV
is the LHC hadronic centre of mass energy, for the relevant ATLAS
constraints we consider, $f_{q,\bar{q}}(x)$ are the parton
distribution functions, and $\tau=M^2/s$.  The sum over quarks is
implicit. We include a $K$-factor to account for next-to-leading-order
corrections, which we take as $K=1.5$  for the purposes of our
analysis.

The parton level cross section for the process, which proceeds
via  $s$-channel exchange of $\gamma$, $Z$, or $Z'$, is given by 
\bea
	&&\hat{\sigma}(q\bar{q}\to l^+l^-)\nonumber\\
 &=& {1\over 32\pi} 
	   {\hat{s}\over (\hat{s}-m_{Z'}^2)^2+\Gamma_{Z'}^2m_{Z'}^2} 
           (v_{q,\sZp}^2+a_{q,\sZp}^2)(v_{l,\sZp}^2+a_{l,\sZp}^2)\nonumber \\
       &+& {1\over 16\pi}{(\hat{s}-m_{Z'}^2)\over (\hat{s}-m_{Z'}^2)^2+\Gamma_{Z'}^2m_{Z'}^2} \times \nonumber \\
        &&\!\!\!\!\!\!\!\!\!\!\!\!\!\ \biggl( {\hat{s}(\hat{s}-m_{Z}^2)\over (\hat{s}-m_{Z}^2)^2+\Gamma_{Z}^2m_{Z}^2}
(v_{q,\sZ} v_{q,\sZp} + a_{q,\sZ} a_{q,\sZp})(v_{l,\sZ} v_{l,\sZp} + a_{l,\sZ} a_{l,\sZp}) \biggr. \nonumber \\
       &-& \biggl. 4 e^2 Q_q v_{q,\sZp} v_{l,\sZp}  \biggr) \nonumber \\
	&+& {1\over 2\pi} \left(  {e^4 Q_q^2 \over \hat{s}} - {e^2 Q_q \over 2 }{\hat{s}-m_{Z}^2\over (\hat{s}-m_{Z}^2)^2+\Gamma_{Z}^2m_{Z}^2} v_{l,\sZ} a_{l,\sZ} \right. \nonumber \\
        &+& \left. {1\over 16}{\hat{s} \over
 (\hat{s}-m_{Z}^2)^2+\Gamma_{Z}^2m_{Z}^2} (v_{q,\sZ}^2 + a_{q,\sZ}^2)(v_{l,\sZ}^2 + a_{l,\sZ}^2)\right )
\eea

The couplings of the $Z$ and $Z'$ to SM fermions, $v_{f,X}$ and $a_{f,X}$,
are as given in eq. \ref{ffcouplings}.
The $Z'$ width, $\Gamma_{Z'}$ is taken to be the decay width to
SM particles, as given by eq. \ref{SMwidth}. We determine the branching ratio to
leptons, using the partial width
\be
	\Gamma(Z' \to l^+l^-) = {5 \alpha \epsilon^2 M_{Z'} \over 24 c_W^2}
\ee
where $l=e$ or $\mu$.

We determine the quantity $\sigma BR(Z' \to l^+l^-)$ as a 
function of the $Z'$ mass, for several choices of the
kinetic mixing parameter, $\epsilon$. 
Our result is shown in fig. \ref{fig:dilepton}. 
From this constraint, we further determine an upper limit on $\epsilon$ as a function of $m_{Z'}$,
equating our predicted cross section to the expected ATLAS limit, in the
combined channel $e^+ e^- + \mu^+\mu^-$. 
The result is shown in fig. \ref{fig:eps-lim}.

\end{document}